\documentclass[a4paper,11pt]{article}
\pdfoutput=1

\usepackage{amsthm,amsmath,amssymb,amsfonts,tikz}
\usepackage{color}
\usepackage[bookmarksnumbered=true]{hyperref}

\usepackage{setspace}
\usetikzlibrary{arrows,positioning,cd,calc}
\usetikzlibrary{decorations.pathreplacing,decorations.markings}
\tikzset{
	on each segment/.style={
		decorate,
		decoration={
			show path construction,
			moveto code={},
			lineto code={
				\path [#1]
				(\tikzinputsegmentfirst) -- (\tikzinputsegmentlast);
			},
			curveto code={
				\path [#1] (\tikzinputsegmentfirst)
				.. controls
				(\tikzinputsegmentsupporta) and (\tikzinputsegmentsupportb)
				..
				(\tikzinputsegmentlast);
			},
			closepath code={
				\path [#1]
				(\tikzinputsegmentfirst) -- (\tikzinputsegmentlast);
			},
		},
	},
	mid arrow/.style={postaction={decorate,decoration={
				markings,
				mark=at position .5 with {\arrow[#1]{stealth}}
	}}},
}

\def\be{\begin{equation}\begin{gathered}}
\def\ee{\end{gathered}\end{equation}}
\newcommand{\rf}[1]{(\ref{#1})}

\numberwithin{equation}{section}

\newtheorem{thm}{Theorem}[section]
\newtheorem{prop}{Proposition}[section]

\newtheorem{conj}{Conjecture}[section]

\theoremstyle{definition}
\newtheorem{Remark}{Remark}[section]

\title{Cluster integrable systems, $q$-Painlev\'e equations\\ and their quantization}
\author{M. Bershtein, P. Gavrylenko, A. Marshakov}

\date{}

\begin{document}

\maketitle

\begin{abstract}\vspace*{2pt}
\noindent
We discuss the relation between the cluster integrable systems and $q$-difference Painlev\'e equations. The Newton polygons corresponding to these integrable systems are all 16 convex polygons with a single interior point. The Painlev\'e dynamics is interpreted as deautonomization of the discrete flows, generated by a sequence of the cluster quiver mutations, supplemented by permutations of quiver vertices.

We also define quantum $q$-Painlev\'e systems by quantization of the corresponding cluster variety. We present formal solution of these equations for the case of pure gauge theory using $q$-deformed conformal blocks or 5-dimensional Nekrasov functions. We propose, that quantum cluster structure of the Painlev\'e system provides
generalization of the isomonodromy/CFT correspondence for arbitrary central charge.
\end{abstract}

\tableofcontents

\newpage

\section{Introduction}

This paper is devoted to certain extension of the relations between four a priori different topics in modern mathematical physics: supersymmetric gauge theories, integrable systems, two-dimensional conformal field theories (CFT), and isomonodromic deformations of differential equations. In fact there are many such relations, but let us distinguish three, mostly important for this work. Remember first that the light (Seiberg-Witten) sector of $\mathcal{N}=2$ supersymmetric 4d gauge theories \cite{SW} can be reformulated in terms of integrable systems \cite{GKMMM}. Second, the Nekrasov partition functions for these theories are equal to the conformal blocks in 2d CFT \cite{AGT}.  Finally, the isomonodromic tau-functions are expressed as the series of conformal blocks (or Nekrasov functions) with special, integer central charge, see \cite{GIL1207}, \cite{ILT}, \cite{GIso}.

These objects and relations among them exist also when have been raised from original setup to ``5d -- relativistic -- $q$-deformed'' framework, moreover the objects and relations acquire some new and nice properties. On the CFT side the conformal symmetry becomes $q$-deformed, and the $q$-deformed W-algebras do have unified description by generators and relations as a quotient of certain quantum group -- the Ding-Iohara-Miki algebra (quantum toroidal ${\mathfrak{gl}}(1)$) \cite{FHSSY:2010}. This algebra has additional properties like $\widetilde{SL}(2,\mathbb{Z})$ action, which are hidden in conformal limit.

On the gauge theory side, the 5d (more precisely 4 plus 1 compact dimension) Nekrasov partition functions are more closely related to the topological strings partition functions, see \cite{Igbal:2003},~\cite{Eguchi:2003}. Integrable systems, corresponding to 5d gauge theories becomes ``relativistic'', see \cite{Nek5d}. This relativization can be understood as passing to a Lie group from its Lie algebra, and it can be more generally formulated in terms of cluster integrable systems \cite{GK:2011},\cite{FM:2014}. On the side of isomonodromic deformations (at least in the first nontrivial example actually used throughout this paper) one has more difference Painlev\'e equations than differential ones, see classification in \cite{SakaiCMP}, \cite{KNY} and similar phenomenon takes place at the level of supersymmetric gauge theories.

As a combination of these relations one gets a link from integrable systems to isomonodromic deformations (or its special case -- the Painlev\'e equations) through gauge theories and CFT. In addition here we propose another, more direct relation. We consider all cluster integrable systems with two-dimensional phase space, and show that their deautonomizations are $q$-difference Painlev\'e equations (all except two), see Theorem \ref{th:qP}. The Poisson structure on the phase space of a cluster integrable system is encoded by certain quiver. Deautonomization of this system is done by switching off one of the basic constraints in their construction, and after that mutations of the quiver (supplemented by permutations of its vertices) start to generate the $q$-Painlev\'e dynamics, as well as the automorphisms of the system. We have constructed explicitly these groups in Sect.~\ref{ss:clqP}.

In the simplest nontrivial example the cluster integrable system is two-particle affine relativistic Toda chain, its deautonomization leads to the $q$-Painlev\'e equation with the surface type $A_7^{(1)'}$. Recall the meaning of other relations in this case: the spectral curve of relativistic Toda is the Seiberg-Witten curve of pure $SU(2)$ 5d theory, the corresponding Nekrasov partition function equals to the Whittaker limit of conformal block for $q$-deformed Virasoro algebra. Finally, it has been proposed in \cite{BS:2016:1}, that certain sum of these Nekrasov 5d partition functions for pure theory in case $q_1q_2=1$ ($c=1$ in CFT terms) is equal to the tau-function of the $q$-Painlev\'e equation $A_7^{(1)'}$, the same equation which we get by deautonomization. This is our main example, some of the statements in Sections \ref{ss:clqP}, \ref{ss:quant} were checked explicitly only in this case.

The fact that the Painlev\'e equations can be obtained by  deautonomization and their relation to integrable systems is not new, see \cite{KMNOY} and other references below, but the cluster nature of the integrable systems corresponding to the $q$-deformed situation has not yet been discussed. In the cluster language the deautonomization arises naturally and unambiguously, and this is one of the advantages of the cluster interpretation.

However, the main advantage of the cluster approach is \emph{quantization}. There is a straightforward way to quantize a classical cluster integrable system, just since it can be formulated on a Poisson variety with almost canonical (logarithmically constant) bracket. In Sect.~\ref{ss:quant} we consider the simplest example it detail. We obtain the quantum Painlev\'e equation just as deautonomization of quantum relativistic affine Toda chain. Then we present a formal solution for the tau-functions of these quantum equations --  as a linear combination of Nekrasov partition functions with generic $\epsilon$-background parameters $(q_1,q_2)$ (arbitrary central charge in CFT terms or the case of refined topological string theory), see Conjecture~\ref{th:tauconf}. The product $p=q_1q_2$ plays the role of multiplicative quantization parameter.
We certainly expect a similar picture for the other $q$-Painlev\'e equations (and probably more generic $q$-Schlesinger systems) --- their quantum version should have formal solutions in terms of Nekrasov partition functions with generic $(q_1,q_2)$.

For the calculations with quivers we used Sage package \cite{sagemath} and Mathematica code from \cite{Caorsi:2017}.
	
\section{Newton polygons and cluster integrable systems
	\label{ss:cluster}}

\subsection{Integrable systems on cluster varieties}
\label{ssec:GKFM}

A \textit{lattice} polygon $\Delta$ is a polygon in the plane $\mathbb{R}^2$ with all vertices in  $\mathbb{Z}^2\subset\mathbb{R}^2$. There is an action of the group $SA(2,\mathbb{Z})=SL(2,\mathbb{Z})\ltimes \mathbb{Z}^2$
on the set of such polygons, which preserves the area, the number of interior points, and the discrete lengths of sides
(number of points on side including vertices minus 1).

Any convex polygon $\Delta$ can be considered as a Newton polygon of polynomial $f_\Delta(\lambda,\mu)$, and equation
\be
\label{SC}
f_\Delta(\lambda,\mu) = \sum_{(a,b)\in\Delta}\lambda^a\mu^b f_{a,b}=0.
\ee
defines a plane (noncompact) spectral curve in $\mathbb{C}^\times\times \mathbb{C}^\times$. The genus $g$
of this curve is equal in general position to the number of integral points strictly inside the polygon $\Delta$.

According to \cite{GK:2011},\cite{FM:2014} a convex Newton polygon $\Delta$ defines a cluster integrable system. First ingredient is an $X$-cluster Poisson variety $\mathcal{X}$, of dimension $\dim_\mathcal{X} = 2S$, where $S$ is an area of the polygon $\Delta$. The Poisson structure in cluster variables is logarithmically constant, and can be encoded by the quiver $\mathcal{Q}$ with $2S$ vertices. Let  $\epsilon_{ij}$ be the number of arrows from $i$-th to $j$-th vertex ($\epsilon_{ji} = -\epsilon_{ij}$) of $\mathcal{Q}$, then Poisson bracket has the form
\be
\label{PB}
\{ y_i, y_j\} = \epsilon_{ij}y_iy_j,\ \ \ \ \{y_i\}\in \left(\mathbb{C}^\times\right)^{2S}.
\ee
The product of all cluster variables $\prod_i y_i$ is a Casimir for the Poisson bracket \eqref{PB}.\footnote{This property means that for each vertex the numbers of outgoing and incoming edges coincide. This is always true for quivers, corresponding to the cluster integrable systems, constructed from dimer models on a torus, or to the integrable systems on the Poisson submanifolds in affine Lie groups.} The phase space of the cluster integrable system is a subvariety of $\mathcal{X}$
of dimension $2S-1$ given by the equation
\be
\label{eq:q}
q=\prod_i y_i=1.
\ee
Integrals of motion for this system are constructed from the coefficients of the spectral curve equation
\rf{SC}, which become functions on $\mathcal{X}$, i.e. $\{f_{a,b}\}=\{f_{a,b}(y)\}$.
More precisely, integrability means
that the coefficients, corresponding to the boundary points of $I\in\bar{\Delta}$, are Casimir functions for the Poisson bracket \rf{PB} (their total number is $B-3$, since the equation \rf{SC} is defined modulo multiplicative renormalization of spectral parameters $\lambda$, $\mu$ and $f_\Delta(\lambda,\mu)$ itself), while properly normalized coefficients, corresponding to the internal points, are integrals of motion
\be
\label{Pinv}
\{ f_{a,b}(y),f_{c,d}(y) \} = 0,\ \ \ \ \ (a,b), (c,d)\in\Delta
\ee
i.e. in involution w.r.t. \rf{PB}. By Pick theorem we have
\be
\label{Pick}
2S-1= (B-3)+2g,
\ee
where $g$ is the number of internal points (or genus of the curve \rf{SC}), or the number of independent integrals of motion.

There are two \emph{a priori} different, but equivalent procedures
of deriving equation \rf{SC}, playing the role of a spectral curve for an integrable system.
\begin{itemize}
	\item A combinatorial way due to \cite{GK:2011} assigns to a Newton polygon a bipartite graph $\Gamma$ on a torus, which can be constructed from a Thurston diagram in the fundamental domain $\mathbb{R}^2/(\mathbb{Z}\times \mathbb{Z})$. The cluster variables $\{y_i\}$ are then assigned to the faces of $\Gamma$, being monodromies of connection around faces of $\Gamma$ (which can be naturally oriented since the graph $\Gamma$ is bipartite).
	
	Equation \rf{SC} is then
vanishing of the dimer partition function on $\Gamma$, computed with some extra sign factors. The weight of any dimer configuration is given by product of Abelian gauge connections ($\mathbb{C}^\times$-valued function) over the marked edges, so that for difference of any two configurations (considered on a bipartite graph $\Gamma$ as elements of 1-chains) this gives an element of $H^1(\Gamma)$. Any such element can be further decomposed into the product $\lambda^a\mu^b f_{a,b}(y)$ from \rf{SC},
where spectral parameters $\lambda,\mu$ are monodromies of the gauge connection or elements of the first cohomology group of the torus $H^1(T^2)\subset H^1(\Gamma)$.  The Poisson structure \rf{PB} is induced by the intersection form in
$H_1(\Sigma)$, where $\Sigma$ is a surface, determined by the bipartite graph $\Gamma$ with dual
fat-graph structure (obtained by twisting of original one at the vertices of one fixed color).
The intersection form on torus itself defines the ``dual two-form'' $\frac{\delta\lambda}{\lambda}\wedge
\frac{\delta\mu}{\mu}$, which can be thought of as variation of the Seiberg-Witten differential and gives
rise to the same symplectic form of the integrable system along the lines of \cite{KricheverPhong}.
	
	\item An equivalent way due to \cite{FM:2014} says that Thurston diagram also defines a word $u\in (W\times W)^\sharp$ in the coextended double affine Weyl group of the affine group $\widehat{PGL}(N)$. For such cyclically irreducible word one can further construct a group element $g(\lambda;y)\in \mathcal{X}_u$ 
	on a Poisson submanifold $\mathcal{X}_u\subset\widehat{PGL}(N)/{\rm Ad}H$, so that the characteristic polynomial
	\be
	\label{CP}
	\det_{N\times N}\left(g(\lambda;y)+\mu\right) = 0
	\ee
	gives the spectral curve equation \rf{SC}. The Poisson structure on submanifold $\mathcal{X}_u$ is given by restriction of the $r$-matrix bracket, and the integrals of motion, computed from \rf{CP} are just ${\rm Ad}H$-invariant functions  \cite{FM:1997,AMJGP,FM:2014}. Formula \rf{CP} provides the most technically effective way for writing the spectral curve equation \rf{SC}, and provides expression for the integrals of motion in terms of cluster variables \cite{AMJGP,KM}, which can be even sometimes generalized to other series beyond $\widehat{PGL}(N)$.
\end{itemize}

\subsection{Polygons with single interior point}

\begin{figure}[h]
	\begin{center}
		\begin{tabular}{c c c c c c c c}
			3 & 4a & 4b & 4c & 5a & 5b & 6a & 6b \vspace{0.2cm}\\
			\begin{tikzpicture}[x=1.5em, y=1.5em, font = \small]
			\draw[fill] (0,1) circle (1pt) -- (1,0) circle (1pt) -- (-1,-1) circle (1pt) -- (0,1);
			\draw[fill] (0,0) circle (1pt);
			\end{tikzpicture}
			&
			\begin{tikzpicture}[x=1.5em, y=1.5em, font = \small]
			\draw[fill] (0,1) circle (1pt) -- (1,0) circle (1pt) -- (0,-1) circle (1pt) -- (-1,0) circle (1pt) -- (0,1);
			\draw[fill] (0,0) circle (1pt);
			\end{tikzpicture}
			&
			\begin{tikzpicture}[x=1.5em, y=1.5em, font = \small]
			\draw[fill] (0,1) circle (1pt) -- (1,0) circle (1pt) -- (-1,-1) circle (1pt) -- (-1,0) circle (1pt) -- (0,1);
			\draw[fill] (0,0) circle (1pt);
			\end{tikzpicture}
			&
			\begin{tikzpicture}[x=1.5em, y=1.5em, font = \small]
			\draw[fill] (0,1) circle (1pt) -- (1,0) circle (1pt) -- (-2,-1) circle (1pt) -- (-1,0) circle (1pt) -- (0,1);
			\draw[fill] (0,0) circle (1pt);
			\end{tikzpicture}
			&
			\begin{tikzpicture}[x=1.5em, y=1.5em, font = \small]
			\draw[fill] (0,1) circle (1pt) -- (1,1) circle (1pt) -- (1,0) circle (1pt) -- (0,-1) circle (1pt) -- (-1,0) circle (1pt) -- (0,1);
			\draw[fill] (0,0) circle (1pt);
			\end{tikzpicture}
			&
			\begin{tikzpicture}[x=1.5em, y=1.5em, font = \small]
			\draw[fill] (0,1) circle (1pt) -- (1,0) circle (1pt) -- (-1,-1) circle (1pt) -- (-1,0) circle (1pt) -- (-1,1) circle (1pt) -- (0,1);
			\draw[fill] (0,0) circle (1pt);	
			\end{tikzpicture}
			&
			\begin{tikzpicture}[x=1.5em, y=1.5em, font = \small]
			\draw[fill] (0,1) circle (1pt) -- (1,1) circle (1pt) -- (1,0) circle (1pt) -- (0,-1) circle (1pt) -- (-1,-1) circle (1pt) -- (-1,0) circle (1pt) -- (0,1);
			\draw[fill] (0,0) circle (1pt);
			\end{tikzpicture}
			&
			\begin{tikzpicture}[x=1.5em, y=1.5em, font = \small]
			\draw[fill] (0,1) circle (1pt) -- (1,1) circle (1pt) -- (1,0) circle (1pt) -- (-1,-1) circle (1pt) -- (-1,0) circle (1pt) -- (-1,1) circle (1pt) -- (0,1);
			\draw[fill] (0,0) circle (1pt);
			\end{tikzpicture} \vspace{0.5cm}\\
			6c & 6d & 7a & 7b & 8a & 8b & 8c & 9 \\
			\begin{tikzpicture}[x=1.5em, y=1.5em, font = \small]
			\draw[fill] (0,1) circle (1pt) -- (1,1) circle (1pt) -- (1,0) circle (1pt) -- (0,-1) circle (1pt) -- (-1,0) circle (1pt) -- (-1,1) circle (1pt)--  (0,1);
			\draw[fill] (0,0) circle (1pt);
			\end{tikzpicture}
			&
			\begin{tikzpicture}[x=1.5em, y=1.5em, font = \small]
			\draw[fill] (0,1) circle (1pt) -- (1,0) circle (1pt) -- (-1,-1) circle (1pt) -- (-1,0) circle (1pt) -- (-1,1) circle (1pt) -- (-1,2) circle (1pt)  -- (0,1);
			\draw[fill] (0,0) circle (1pt);
			\end{tikzpicture}
			&
			\begin{tikzpicture}[x=1.5em, y=1.5em, font = \small]
			\draw[fill] (0,1) circle (1pt) -- (1,0) circle (1pt) -- (0,-1) circle (1pt) -- (-1,-1) circle (1pt) -- (-1,0) circle (1pt) -- (-1,1) circle (1pt) -- (-1,2) circle (1pt)  -- (0,1);
			\draw[fill] (0,0) circle (1pt);
			\end{tikzpicture}
			&
			\begin{tikzpicture}[x=1.5em, y=1.5em, font = \small]
			\draw[fill] (0,1) circle (1pt) -- (1,0) circle (1pt) -- (1,-1) circle (1pt) -- (0,-1) circle (1pt) -- (-1,-1) circle (1pt) -- (-1,0) circle (1pt) -- (-1,1) circle (1pt) --  (0,1);
			\draw[fill] (0,0) circle (1pt);
			\end{tikzpicture}
			&
			\begin{tikzpicture}[x=1.5em, y=1.5em, font = \small]
			\draw[fill] (0,1) circle (1pt) -- (1,-1) circle (1pt) -- (0,-1) circle (1pt) -- (-1,-1) circle (1pt) -- (-1,0) circle (1pt) -- (-1,1) circle (1pt) -- (-1,2) circle (1pt) -- (-1,3) circle (1pt) -- (0,1);
			\draw[fill] (0,0) circle (1pt);	
			\end{tikzpicture}
			&
			\begin{tikzpicture}[x=1.5em, y=1.5em, font = \small]
			\draw[fill] (0,1) circle (1pt) -- (1,0) circle (1pt) -- (1,-1) circle (1pt) -- (0,-1) circle (1pt) -- (-1,-1) circle (1pt) -- (-1,0) circle (1pt) -- (-1,1) circle (1pt) -- (-1,2) circle (1pt) -- (0,1);
			\draw[fill] (0,0) circle (1pt);	
			\end{tikzpicture}
			&
			\begin{tikzpicture}[x=1.5em, y=1.5em, font = \small]
			\draw[fill] (0,1) circle (1pt) -- (1,1) circle (1pt) -- (1,0) circle (1pt) -- (1,-1) circle (1pt) -- (0,-1) circle (1pt) -- (-1,-1) circle (1pt) -- (-1,0) circle (1pt) -- (-1,1) circle (1pt) -- (0,1);
			\draw[fill] (0,0) circle (1pt);	
			\end{tikzpicture}	
			&
			\begin{tikzpicture}[x=1.5em, y=1.5em, font = \small]
			\draw[fill] (0,1) circle (1pt) -- (1,0) circle (1pt) -- (2,-1) circle (1pt) -- (1,-1) circle (1pt) -- (0,-1) circle (1pt) -- (-1,-1) circle (1pt) -- (-1,0) circle (1pt) -- (-1,1) circle (1pt) -- (-1,2) circle (1pt) -- (0,1);
			\draw[fill] [](0,0) circle (1pt);	
			\end{tikzpicture}
		\end{tabular}
		\caption{List of Newton polygons with a single internal point and $3\leq B\leq 9$ boundary points.}
		\label{fi:NP}
	\end{center}
\end{figure}
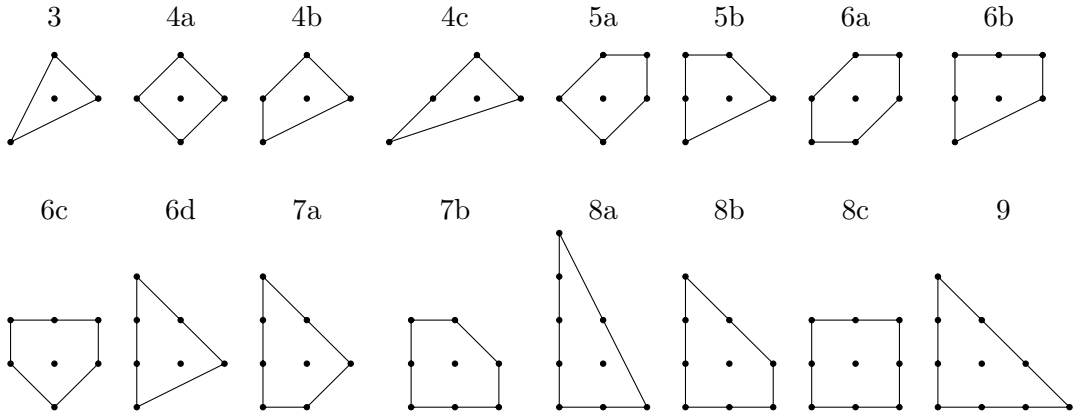

It is well-known that any  convex lattice polygon with the only lattice
point in the interior is equivalent by $SA(2,\mathbb{Z})$ to one of the  16 polygons from Fig~\ref{fi:NP}.
We label them $Bx$, where $B$ is a number of the boundary points, and letter $x$ distinguished their types,
if there are several for given $B$. Since there is only one interior point, the genus of the corresponding spectral curve \eqref{SC} is just $g=1$, and one finds from \rf{Pick} for the area  $S=B/2$.


Using the methods of \cite{GK:2011,FM:2014} briefly described above, one can reconstruct the Thurston diagrams, bipartite graphs, and quivers, corresponding to the polygons from the fig.~\ref{fi:NP}, which are collected  together with the bipartite graphs in Appendix~\ref{ap:catalog}. On the other side, by \cite[Theorem 3.12]{GK:2011}, one can reconstruct Newton polygon from dimer partition function of the bipartite graph. This correspondence between polygons with single interior point, bipartite graphs, and quivers is not new, see \cite{HS}, \cite{Franco:2017} and references therein.

\begin{figure}[h]
\begin{center}
\begin{tabular}{c c c c c c c}
	$A_8^{(1)}$ & $A_7^{(1)'}$ & $A_7^{(1)}$ & $A_6^{(1)}$ & $A_5^{(1)}$ &  \vspace{0.3cm}
	\\
	\begin{tikzpicture}[scale=1.5, font = \small]
\node[shape=circle,draw=black] (C) at (-0.5,0) {\color{blue}3};
\node[shape=circle,draw=black] (B) at (0.5,0) {\color{blue}2};
\node[shape=circle,draw=black] (A) at (0,cos{30}) {\color{blue}1};
\path
\foreach \from/\to in {A/B,B/C,C/A}
{
	(\from) edge[mid arrow,bend right=20] (\to)
	(\from) edge[mid arrow,bend left=20] (\to)
	(\from) edge[mid arrow] (\to)
};			
\end{tikzpicture} 			
	&
\begin{tikzpicture}[scale=1.5, font = \small]
\node[shape=circle,draw=black] (A) at (0,1) {\color{blue}1};
\node[shape=circle,draw=black] (B) at (1,1) {\color{blue}2};
\node[shape=circle,draw=black] (C) at (1,0) {\color{blue}3};
\node[shape=circle,draw=black] (D) at (0,0) {\color{blue}4};
\path \foreach \from/\to in {A/B,B/C,C/D,D/A}
{
	(\from) edge[mid arrow,bend right=15] (\to)
	(\from) edge[mid arrow,bend left=15] (\to)
};			
\end{tikzpicture}			
	&
\begin{tikzpicture}[scale=1.5, font = \small]
\node[shape=circle,draw=black] (A) at (0,1) {\color{blue}1};
\node[shape=circle,draw=black] (B) at (1,1) {\color{blue}2};
\node[shape=circle,draw=black] (C) at (1,0) {\color{blue}3};
\node[shape=circle,draw=black] (D) at (0,0) {\color{blue}4};
\path
\foreach \from/\to in {A/B,C/D,D/A}
{
	(\from) edge[mid arrow,bend right=20] (\to)
	(\from) edge[mid arrow,bend left=20] (\to)
};			
\path
\foreach \from/\to in {B/C,B/D,D/A,A/C}
{
	(\from) edge[mid arrow] (\to)
};			
\end{tikzpicture}
	&
\begin{tikzpicture}[scale=1.5, font = \small]
\node[shape=circle,draw=black] (A) at (0,1.62) {\color{blue}2};
\node[shape=circle,draw=black] (B) at (0.8,0.95) {\color{blue}3};
\node[shape=circle,draw=black] (C) at (0.5,0) {\color{blue}4};
\node[shape=circle,draw=black] (D) at (-0.5,0) {\color{blue}5};
\node[shape=circle,draw=black] (E) at (-0.8,0.95) {\color{blue}1};
\path
\foreach \from/\to in {A/B,E/A}
{
	(\from) edge[mid arrow,bend right=15] (\to)
	(\from) edge[mid arrow,bend left=15] (\to)
}
\foreach \from/\to in {B/C,B/D,C/E,D/E,C/D,A/C,D/A}
{
	(\from) edge[mid arrow] (\to)					
}
;				
\end{tikzpicture} 						
	&
\begin{tikzpicture}[scale=1.5, font = \small]
\node[shape=circle,draw=black] (A) at (0,1.73) {\color{blue}1};
\node[shape=circle,draw=black] (B) at (1,1.73) {\color{blue}2};
\node[shape=circle,draw=black] (C) at (1.5,0.865) {\color{blue}3};
\node[shape=circle,draw=black] (D) at (1,0) {\color{blue}4};
\node[shape=circle,draw=black] (E) at (0,0) {\color{blue}5};
\node[shape=circle,draw=black] (F) at (-0.5,0.865) {\color{blue}6};
\path
\foreach \from/\to in {A/B,B/C,C/D,D/E,E/F,F/A,A/C,C/E,E/A,B/D,D/F,F/B}
{
	(\from) edge[mid arrow] (\to)					
}
;				
\end{tikzpicture} 						
\end{tabular}

\medskip

\begin{tabular}{c c c}
	$A_4^{(1)}$ & $A_3^{(1)}$ & $A_2^{(1)}$ \vspace{0.2cm}
	\\
\begin{tikzpicture}[scale=2, font = \small]
\node[shape=circle,draw=black] (A) at (0,1) {\color{blue}1};
\node[shape=circle,draw=black] (B) at (-0.4,1.4) {\color{blue}2};
\node[shape=circle,draw=black] (C) at (1,1) {\color{blue}3};
\node[shape=circle,draw=black] (D) at (1.4,1.4) {\color{blue}4};
\node[shape=circle,draw=black] (E) at (1,-0.3) {\color{blue}5};
\node[shape=circle,draw=black] (F) at (0.5,0.0) {\color{blue}6};
\node[shape=circle,draw=black] (G) at (0,-0.3) {\color{blue}7};
\path
\foreach \from/\to in {B/D,D/E,E/G,G/A,D/F,G/B,F/B}
{
	(\from) edge[mid arrow, bend left=20] (\to)					
}
\foreach \from/\to in {A/D,B/C,E/F,F/G}
{
	(\from) edge[mid arrow] (\to)					
}
\foreach \from/\to in {A/C,C/E,C/F,F/A}
{
	(\from) edge[mid arrow, bend right=20] (\to)					
}
;		
\end{tikzpicture}& 						
\begin{tikzpicture}[scale=2, font = \small]
\node[shape=circle,draw=black] (A) at (0,1) {\color{blue}1};
\node[shape=circle,draw=black] (A1) at(-0.3,1.3) {\color{blue}2};
\node[shape=circle,draw=black] (B) at (1,1) {\color{blue}3};
\node[shape=circle,draw=black] (B1) at(1.3,1.3) {\color{blue}4};
\node[shape=circle,draw=black] (C) at (1,0) {\color{blue}5};
\node[shape=circle,draw=black] (C1) at(1.3,-0.3) {\color{blue}6};
\node[shape=circle,draw=black] (D) at (0,0) {\color{blue}7};
\node[shape=circle,draw=black] (D1) at(-0.3,-0.3) {\color{blue}8};
\path
\foreach \from/\to in {A/B1,A1/B,B/C1,B1/C,C/D1,C1/D,D/A1,D1/A}
{
	(\from) edge[mid arrow] (\to)					
}
\foreach \from/\to in {A/B,B/C,C/D,D/A}
{
	(\from) edge[mid arrow,bend right=15] (\to)					
}
\foreach \from/\to in {A1/B1,B1/C1,C1/D1,D1/A1}
{
	(\from) edge[mid arrow,bend left=15] (\to)					
}
;
\end{tikzpicture}
	&
\begin{tikzpicture}[scale=2, font = \small]
\node[shape=circle,draw=black] (A) at (-0.5,0) {\color{blue}1};
\node[shape=circle,draw=black] (A1) at (-0.76,-0.15) {\color{blue}2};
\node[shape=circle,draw=black] (A2) at (-1.02,-0.3) {\color{blue}3};	
\node[shape=circle,draw=black] (B) at (0.5,0) {\color{blue}7};
\node[shape=circle,draw=black] (B1) at (0.76,-0.15) {\color{blue}8};
\node[shape=circle,draw=black] (B2) at (1.02,-0.30) {\color{blue}9};	
\node[shape=circle,draw=black]  (C) at (0,0.86) {\color{blue}4};
\node[shape=circle,draw=black] (C1) at (0,1.16) {\color{blue}5};
\node[shape=circle,draw=black] (C2) at (0,1.46) {\color{blue}6};	
\path
\foreach \from/\to in {C1/B,B1/A,A1/C}
{
	(\from) edge[mid arrow] (\to)					
}
\foreach \from/\to in {C2/B,C2/B1,C2/B2,B2/A,B2/A1,B2/A2,A2/C,A2/C1,A2/C2}
{
	(\from) edge[mid arrow,bend left=30] (\to)					
}
\foreach \from/\to in {C/B,B/A,A/C}
{
	(\from) edge[mid arrow,bend right=30] (\to)					
}
\foreach \from/\to in {C/B1,B/A1,A/C1}
{
	(\from) edge[mid arrow,bend right=20] (\to)					
}
\foreach \from/\to in {C/B2,B/A2,A/C2}
{
	(\from) edge[mid arrow,bend right=10] (\to)					
}
\foreach \from/\to in {C1/B2,B1/A2,A1/C2}
{
	(\from) edge[mid arrow,bend left=15] (\to)					
}
\foreach \from/\to in {C1/B1,B1/A1,A1/C1}
{
	(\from) edge[mid arrow,bend left=8] (\to)					
}
;
\end{tikzpicture}
\end{tabular}
\caption{Quivers corresponding to polygons with one interior point}
\label{fi:PQ}
\end{center}
\end{figure}
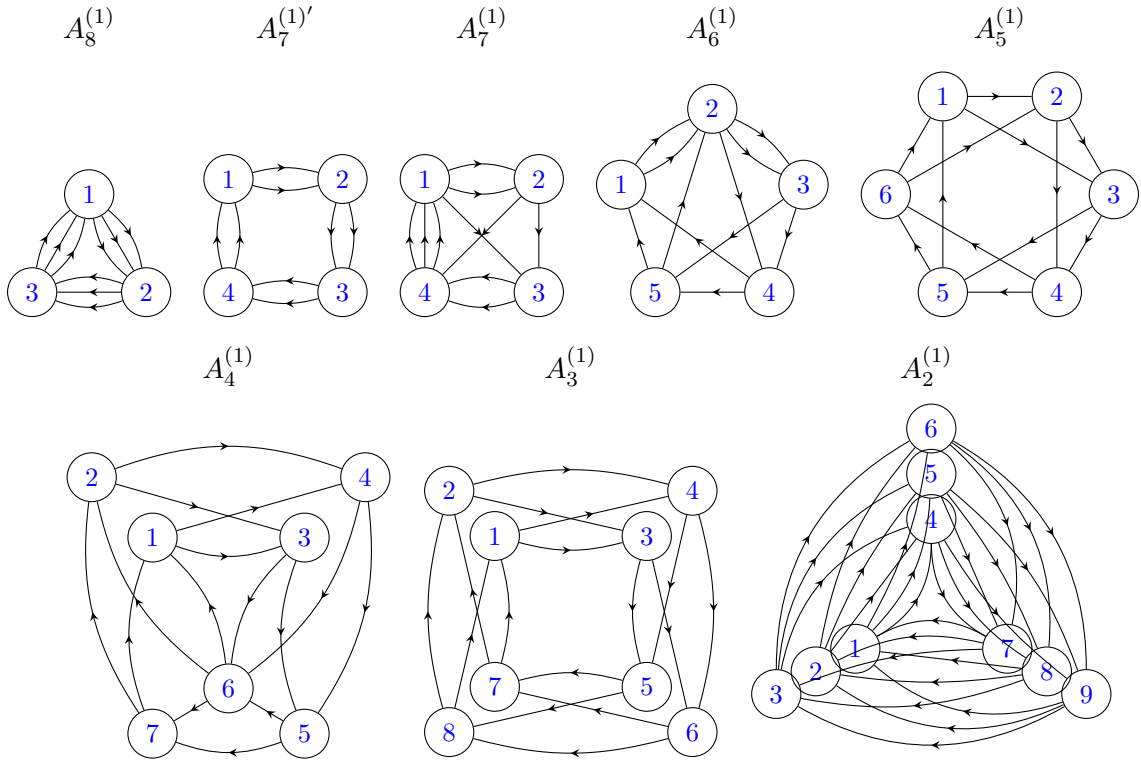

Notice, that from 16 polygons we get only 8 different quivers, presented together at Fig.~\ref{fi:PQ}. We label them by affine root systems for the future relation to $q$-Painlev\'e equations in Sakai's classification.

The correspondence between Fig.~\ref{fi:NP} and Fig.~\ref{fi:PQ} is easily established just by counting the number of the vertices. Generally, a polygon with $B$ boundary points corresponds to the quiver with $B$ vertices, labeled as $A_{11-B}^{(1)}$ at Fig.~\ref{fi:PQ}. The only exceptional case is $B=4$, where the polygons $4a$, $4c$ correspond to the $A_7^{(1)'}$ quiver, and the polygon $4b$ corresponds to different $A_7^{(1)}$ quiver.

\begin{Remark}
The action of $SL(2,\mathbb{Z})$ on polygons $\Delta$ is equivalent to the multiplicative transformation of variables $\lambda=\tilde{\lambda}^a\tilde{\mu}^b$, $\mu=\tilde{\lambda}^c\tilde{\mu}^d$ in the Laurent polynomial $f_{\Delta}(\lambda,\mu)$ for integer $\{a,b,c,d\}\in \mathbb{Z}$, satisfying $ad-bc=1$. One can consider more general rational transformations $\lambda=\lambda(\tilde{\lambda},\tilde{\mu})$, $\mu=\mu(\tilde{\lambda},\tilde{\mu})$, and one can check that if Newton polygons $\Delta_1$, $\Delta_2$ from Fig.~\ref{fi:NP} correspond to the same quiver, then there exist a  rational transformation  which maps equation $f_{\Delta_1}(\lambda,\mu)=0$ to $f_{\Delta_2}(\lambda,\mu)=0$.

For example, take
\be
f_{4a}(\lambda,\mu)=a_{1,0}\lambda+a_{0,1}\mu+a_{0,0}+a_{-1,0}\lambda^{-1}+a_{0,-1}\mu^{-1}
\ee
and substitute there $\lambda=\tilde{\lambda}(a_{-1,0}+a_{0,1}\tilde{\lambda}\tilde{\mu})$, $\mu=\tilde{\mu}/(a_{-1,0}+a_{0,1}\tilde{\lambda}\tilde{\mu})$, then equation for the Newton polygon 4a
\be
f_{4a}(\lambda
,\mu
)=a_{0,1} a_{1,0}\tilde{\lambda}^2\tilde{\mu}+(  a_{0,-1} a_{0,1}+a_{-1,0} a_{1,0})\tilde{\lambda} +a_{0,0}+a_{-1,0}a_{0,-1}\tilde{\mu}^{-1}+\tilde{\lambda}^{-1}=
 f_{4c}(\tilde{\lambda},\tilde{\mu})
\ee
turns in new variables into equation for the polygon 4c.

This is certainly a well-known phenomenon, see, for example discussion around eq.~(3.70) in \cite{Gaiotto}.  However, this is an important issue in our context, and we are going to return to it elsewhere.
\end{Remark}

\subsection{Poisson maps and discrete flows}
\label{ssec:2.3}
There are several classes of the Poisson maps between $X$-cluster varieties. The simplest example is just permutation of the vertices of a quiver, together with the cluster variables $\{y_i\}$ assigned to the vertices, complemented with corresponding permutations of the edges.

Another type of Poisson maps is given by cluster mutations. A mutation can be performed at any vertex. Denote by $\mu_j$ the mutation at $j$-th vertex. It acts as
\be
\label{mutz}
\mu_j:\ \  y_j\mapsto y_j^{-1},\qquad  y_i\mapsto y_i\left(1+y_j^{{\rm sgn}(\epsilon_{ij})}\right)^{\epsilon_{ij}},\ \  i\neq j, 
\ee
supplemented by transformation of the quiver $\mathcal{Q}$ itself, so that
\be
\label{mute}
\epsilon_{ik}\mapsto\epsilon_{ik}+\frac{\epsilon_{ij}|\epsilon_{jk}|+\epsilon_{jk}|\epsilon_{ij}|}{2}.
\ee
Denote by $\mathcal{G}_\mathcal{Q}$ the stabilizer of the quiver $\mathcal{Q}$ --- the group consisting of compositions of mutations and permutations of vertices which map quiver $\mathcal{Q}$ to itself. Such transformations
nevertheless generate nontrivial maps for the cluster variables $\{ y_i\}$. This group is sometimes called mapping class group of $X$-cluster variety.

There is another (but related) group $\mathcal{G}_\Delta$ introduced in \cite{GK:2011},\cite{FM:2014}. It was shown in \cite{GK:2011} that there are certain transformations of bipartite graph, which preserve the dimer partition function. These transformations include elimination of a two-valent vertices and spider
moves. The group $\mathcal{G}_\Delta$ consists of compositions of such transformations which map bipartite graph to itself (also with certain permutations, see below). In terms of transformations of the quivers the spider moves correspond to certain mutations (but not all mutations can be obtained as spider moves), therefore the group $\mathcal{G}_\mathcal{Q}$ is generally larger than
$\mathcal{G}_\Delta$.

Under the condition \eqref{eq:q} the group $\mathcal{G}_\Delta$ can be considered as a group of the discrete flows of cluster integrable system: the integrals of motion are preserved by these discrete flows.
For all our systems, generated by the Newton polygons from Fig.~\ref{fi:NP}, integrability \rf{Pinv} is almost a trivial statement, since here is a single integral of motion $f_{a,b}(y)=H$, for $(a,b)\in\Delta\setminus\bar{\Delta}$.
However, this Hamiltonian is invariant with respect to the discrete flows. Moreover, the (properly normalized) Hamiltonian remains invariant under the action of full $\mathcal{G}_\mathcal{Q}$, which generally acts nontrivially on the Casimir functions.

Let us now list few examples of the Hamiltonians and discrete flows from $\mathcal{G}_\Delta$ for certain quivers from Fig.~\ref{fi:PQ}. The group  $\mathcal{G}_\mathcal{Q}$ will be discussed in the next section, where we also forget about the restriction~\eqref{eq:q}, or deautonomize the cluster integrable system, and the list of the
Hamiltonians, invariant under the action of $\mathcal{G}_\mathcal{Q}$ will be also completed there. Notice, that exact expressions for these Hamiltonians, which are in the Poisson involution \rf{Pinv} (in the nontrivial case
with several internal points) or are singled our being invariant under the action of discrete flow, generally
contain \emph{fractional} powers of the cluster variables \rf{PB} (see also discussion of this point in \cite{FM:2014,AMJGP,KM}).


\paragraph{$\mathbf{A_8^{(1)}}$.} The group $\mathcal{G}_\Delta$ contains just cyclic permutations $\pi=(1,2,3)$ (or rotation on the quiver to $120^\circ$).

Denote $x=y_1$, $y=y_2$, $z=y_3$, then $xyz=1$ and $\{x,y\} = 3xy$, the invariant Hamiltonian is
\begin{equation}
H = x^{2/3}y^{1/3}+ \frac{y^{1/3}}{x^{1/3}} + \frac1{x^{1/3}y^{2/3}}.
\end{equation}

\paragraph{$\mathbf{A_7^{(1)'}}$.} The group $\mathcal{G}_\Delta$ contains
\begin{equation}
\label{Trt1}
\pi_2^2=(1,3)(2,4),\;   T=(1,2)(3,4)\circ \mu_1\circ \mu_3.
\end{equation}
Denote $x=y_1$, $y=y_2$, $Z=y_1y_3$, then $\{x,y\} = 2xy$ and $Z$ is the Casimir function. Transformation	 $T$ acts as $(x,y) \mapsto (y\frac{(x+Z)^2}{(x+1)^2},x^{-1})$. The Hamiltonian, invariant under this transformation, has the form
\begin{equation}
\label{HToda}
H = \sqrt{xy}+ \sqrt{\frac{x}{y}} + \frac1{\sqrt{xy}}
+ Z\sqrt{\frac{y}{x}}
\end{equation}
This is the Hamiltonian of relativistic two-particle affine Toda chain. At $Z\to 0$ it turns into the Hamiltonian of open relativistic Toda chain, first appeared in this context in \cite{FM:1997}.

\paragraph{$\mathbf{A_7^{(1)}}$.} The group $\mathcal{G}_\Delta$ contains element $T=(1324)\circ \mu_3$, it has infinite order and generates subgroup $\mathbb{Z}\subset \mathcal{G}_\Delta$.

Denote $x=y_3$, $y=y_4$, $Z=y_2\sqrt{\frac{y_4}{y_3}}$, then $\{x,y\} = 2xy$ and $Z$ is the Casimir function. The transformation $T$ acts as $(x,y) \mapsto (\frac{1}{Z\sqrt{x^3y}}(1+x),Z\frac{\sqrt{x}}{\sqrt{y}}(1+x))$. The Hamiltonian, invariant under such transformation, has the form
\begin{equation}
H = \sqrt{xy}+ \sqrt{\frac{x}{y}} + \frac1{\sqrt{xy}} + \frac{Z}{x}
\end{equation}
This Hamiltonian is different from the previous one, though it has the same limit at $Z\to 0$, so one can think of this system as of different affinization of two-particle relativistic Toda.

	\section{Cluster mutations and $q$-difference Painlev\'e equations
	\label{ss:clqP}}

\subsection{Poisson cluster varieties and $q$-Painlev\'e equations} \label{ssec:clqPX}

Let us now start from any of the quivers $\mathcal{Q}$ from Fig.~\ref{fi:PQ} and act by elements of the group $\mathcal{G}_\mathcal{Q}$ to the cluster $y$-variables, for generic $q\neq 1$, i.e. forgetting the condition \eqref{eq:q}, necessary for construction of a cluster integrable system.
In this way we obtain $q$-difference Painlev\'e equations as deautonomization of a cluster integrable system, corresponding to the Newton polygons from Fig.~\ref{fi:NP} and quivers from Fig.~\ref{fi:PQ}.
Deautonomization is a standard method to obtain difference Painlev\'e equations, see \cite{Grammaticos} for review and \cite{Dzhamaj} for recent geometric interpretation. However, here we get $q$-Painlev\'e dynamics from deautonomization of the discrete flows in cluster integrable systems, and describe symmetries of the
$q$-difference Painlev\'e equations in terms of the group, generated by cluster mutations and permutations.

It will be convenient also to consider inversion $\varsigma$, the transformation which reverses orientations of all edges and maps all $\varsigma: \{y_i\} \mapsto \{y_i^{-1}\}$. Note, that $\varsigma$ changes the sign of the Poisson structure, this is natural since it reverses the ``time direction''. We denote the corresponding extended group by $\tilde{\mathcal{G}}_\mathcal{Q}$.

\begin{thm} \label{th:qP}
For each quiver from Fig.~\ref{fi:PQ}
the group $\tilde{\mathcal{G}}_\mathcal{Q}$ contains subgroup isomorphic to the symmetry group of the corresponding $q$-Painlev\'e equation and its action on variables $y_i$ is equivalent to $q$-Painlev\'e dynamics.
\end{thm}

Note that the symmetry group of $q$-Painlev\'e equation is the extended affine Weyl group, therefore the Theorem \ref{th:qP} says that $\tilde{\mathcal{G}}_\mathcal{Q}$ contains such subgroup. We believe that actually $\tilde{\mathcal{G}}_\mathcal{Q}$ coincides with this subgroup, see Remark \ref{rem:geom} below.

This theorem is proven by a case-by-case analysis below. Note that this relation between $q$-Painlev\'e equations and cluster mutations (but without relation to cluster integrable systems) was already noticed in \cite{Okubo:2015} for $A_3^{(1)}$ and $A_5^{(1)}$ equations, in \cite{Okubo:2017} for $A_6^{(1)}$ and $A_7^{(1)}$ equations, and mentioned for $A_7^{(1)'}$ equation in \cite{BS:2016:1}. See also \cite{Hone:2014}, where relations to cluster integrable systems were also mentioned.

\begin{figure}[h]
	\begin{center}
		{\small
			\begin{tikzcd}[row sep=scriptsize, column sep=scriptsize]
				& & & & & & & \dfrac{A_7^{(1)}}{A_1^{1},|\alpha^2|=8} \arrow[rd] &\\
				\dfrac{A_0^{(1)}}{E_8^{(1)}} \arrow[r] & \dfrac{A_1^{(1)}}{E_7^{(1)}}  \arrow[r] &	
				\dfrac{A_2^{(1)}}{E_6^{(1)}} \arrow[r] &
				\dfrac{A_3^{(1)}}{E_5^{(1)}}  \arrow[r]  &	
				\dfrac{A_4^{(1)}}{E_4^{(1)}} \arrow[r]  &
				\dfrac{A_5^{(1)}}{E_3^{(1)}}  \arrow[r]  &	
				\dfrac{A_6^{(1)}}{E_2^{(1)}} \arrow[r]  \arrow[ru] & \dfrac{A_7^{(1)'}}{E_1^{(1)}}  &
				\dfrac{A_8^{(1)}}{E_0^{(1)}}
		\end{tikzcd}}
	\end{center}
	\caption{$q$-Painlev\'e equations by surface/symmetry type}
	\label{fi:qPeq}
\end{figure}
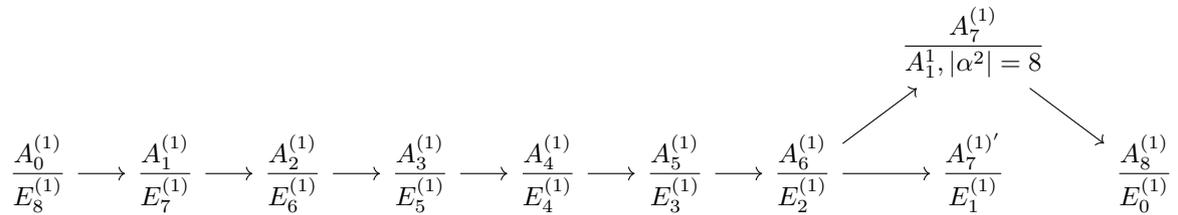

At Fig.~\ref{fi:qPeq} we represent all $q$-Painlev\'e equations following \cite{SakaiCMP}, the arrows stand for degenerations of the equations.

\begin{Remark}\label{rem:geom}
	In the geometric approach to difference Painlev\'e equations the main object of interest is special surface, constructed as blowup of $\mathbb{CP}^2$ in nine points \cite{SakaiCMP}, and commonly called as generalized Halphen or half-K3 surface. It is natural to ask therefore, how it comes out in our approach.
	
	Each seed (quiver with the set of variables $y_i$) corresponds to a chart in Poisson cluster $X$-variety. For quivers from Fig.~\ref{fi:PQ} the rank of the Poisson bracket is always equal to two, or in other words if the values of the Casimir functions are fixed, one gets a surface. We conjecture, that after gluing such surfaces for all charts (or seeds) one obtains the half-K3 surface, corresponding to the Painlev\'e equation (perhaps up to submanifold of codimension two -- a finite set of points).
	
	Note also, that it follows from the results of \cite{SakaiCMP}, that if one takes not all charts (or seeds), but only those -- obtained by the action of the subgroup from Theorem~\ref{th:qP}, one gets the corresponding Painlev\'e surface. The above conjecture is stronger, it deals with all charts, which, in principle, could correspond to further blowups. If this conjecture holds, the group $\tilde{\mathcal{G}}_\mathcal{Q}$ coincides with the symmetry group of the $q$-Painlev\'e equation, since there is no other isomorphisms between such surfaces. 
\end{Remark}


\paragraph{$\mathbf{A_8^{(1)}}$.} The group $\tilde{\mathcal{G}}_\mathcal{Q}$ contains cyclic permutation $\pi=(1,2,3)$ and also $\sigma=(1,2)\circ \varsigma$. This group is the symmetry group of this Painlev\'e equation. 

\paragraph{$\mathbf{A_7^{(1)'}}$.}The group $\tilde{\mathcal{G}}_\mathcal{Q}$ contains
\begin{equation}
\pi_1=(1,3)\circ\varsigma,\;\pi_2=(4,3,2,1),\;   T=(1,2)(3,4)\circ \mu_1\circ \mu_3.
\end{equation}
Applying $T=(1,2)(3,4)\circ \mu_1\circ \mu_3$ we get
\begin{equation}
(y_1, y_2, y_3, y_4)\mapsto \left(y_2\frac{(y_3+1)^2}{(y_1^{-1}+1)^2},y_1^{-1},y_4\frac{(y_1+1)^2}{(y_3^{-1}+1)^2},y_3^{-1}\right).
\end{equation}
Now we substitute
\begin{equation}\label{eq:GA7p}
q=y_1y_2y_3y_4,\quad Z=y_2^{-1}y_4^{-1},\quad F=y_1,\quad G=y_2^{-1}
\end{equation}
and get
\be
\label{Trt2}
\pi_2 \colon (Z,q,F,G)\mapsto \left(Z^{-1}q^{-1},q,Z^{-1}G,F^{-1}\right),
\\
\pi_1 \colon (Z,q,G,F)\mapsto \left(Z^{-1},q^{-1},q^{-1}Z^{-1}F^{-1},G^{-1}\right),
\\
T\colon (Z,q,F,G)\mapsto \left(q Z,q,\frac{(F+q Z)^2}{(F+1)^2 G},F\right).
\ee
These transformations $\pi_1$, $\pi_2$, $T$ generate the symmetry group of the $q$-difference Painlev\'e equation with surface type $A_7^{(1)'}$ as written in \cite{BS:2016:1}.
The relations on them are
\[\pi_2^4=\pi_1^2=(\pi_1\pi_2)^2=e,\quad \pi_2 T \pi_2 =T^{-1}, \quad \pi_1 T\pi_1=\pi_2^2T.\]
If we introduce $s_0=\pi_2 T$ and $s_1=T \pi_2$ then these elements satisfy $s_0^2=s_1^2=e$ and generate subgroup isomorphic to Weyl group of the affine roots system $A_1^{(1)}$. The full group is semidirect product $Dih_4\ltimes W(A^{(1)}_1)$ where $Dih_4$ is dihedral group of square.

Now consider $G,F$ as a functions of $Z,q$ such that $T(G)=G(qZ)$, $T(F)=F(qZ)$. Then the formulas \eqref{Trt2} lead to the equation
\begin{equation}\label{eq:qPAtp}
G(qZ)G(q^{-1}Z)=\frac{(G(Z)+Z)^2}{(G(Z)+1)^2}
\end{equation}
This second order $q$-difference equation is called $q$-Painlev\'e equation (of the type $A_7^{(1)'}$).

\paragraph{$\mathbf{A_7^{(1)}}$.}	Applying $T=(1324)\circ \mu_3$ we get
\begin{equation}
(y_1, y_2, y_3, y_4)\mapsto \left(y_4(1+y_3^{-1})^{-2},y_3^{-1},y_1(1+y_3),y_2(1+y_3)\right).
\end{equation}
Now we substitute
\begin{equation}
q=y_1y_2y_3y_4,\quad Z=y_1y_2^{-1}y_3^{2},\quad F=y_2^{-1}y_3,\quad G=y_3
\end{equation}
and get
\begin{equation}
(Z,q,F,G)\mapsto\left(q Z,q, Z(1+G)F^{-1}, Z(1+G)F^{-1}G^{-1}\right),
\end{equation}
This dynamics is equivalent to $q$-difference Painlev\'e equation with surface type $A_7^{(1)}$ as written in \cite[Appendix C]{SakaiCMP}. One may
also rewrite it as an equation for single function $G(Z)$:
\begin{equation}
G(qZ){G(Z)^2}G(q^{-1}Z)=Z(G(Z)+1)
\end{equation}

\paragraph{$\mathbf{A_6^{(1)}}$.} The group $\tilde{\mathcal{G}}_\mathcal{Q}$ contains elements
\begin{equation}
s_0=(1,3)\circ\mu_1\circ\mu_3,\;\;s_1=(4,5)\circ\mu_2\circ \mu_4\circ\mu_5\circ\mu_2,\;\;\sigma=(1,3)(4,5)\circ\varsigma,\;\;T=(1,2,3,5,4)\circ \mu_5,\;
\end{equation}
The $\sigma$ is just a composition of inversion and permutation and other transformations look like	
\be
s_1\colon (y_1,y_2,y_3,y_4,y_5) \mapsto \left(y_1y_2y_5^2\frac{(1{+}y_4{+}y_2^{-1})(1{+}y_2{+}y_5^{-1})}{(1{+}y_5{+}y_4^{-1})^2},
\frac{y_2}{y_4y_5}\frac{1{+}y_4{+}y_2^{-1}}{1{+}y_2{+}y_5^{-1}},\right.
\\ \quad\quad\quad\quad\quad\quad 
\left.y_2y_3y_4^{2}\frac{(1{+}y_5{+}y_4^{-1})^2}{(1{+}y_2{+}y_5^{-1})(1{+}y_4{+}y_2^{-1})},
\frac{y_4}{y_2y_5}\frac{1{+}y_5{+}y_4^{-1}}{1{+}y_4{+}y_2^{-1}},
\frac{y_5}{y_2y_4}\frac{1{+}y_2{+}y_5^{-1}}{1{+}y_5{+}y_4^{-1}}\right)
\\
s_0\colon (y_1,y_2,y_3,y_4,y_5) \mapsto \left(y_3^{-1},y_2\frac{(1+y_3)^2}{(1+y_1^{-1})^2},y_1^{-1},y_4\frac{1+y_1}{1+y_3^{-1}},
y_5\frac{1+y_1}{1+y_3^{-1}}\right)
\\
T\colon (y_1,y_2,y_3,y_4,y_5) \mapsto \left(y_4(1+y_5),\frac{y_1}{1+y_5^{-1}},\frac{y_2}{1+y_5^{-1}},y_5^{-1},y_3(1+y_5)\right)
\ee
The elements $s_0,s_1$ generate affine Weyl group $W(A_1^{(1)})$. Let $\pi=\sigma \circ T$, then the elements $r_0=\pi\sigma\pi,r_1=\sigma$ generate affine Weyl group $W(A_1^{(1)})$, these two Weyl groups commute. The conjugation by $\pi$ preserves these subgroups, the full group generated by $T,\sigma,s_0,s_1$ is an extended affine Weyl group $\tilde{W}((A_1+A_1)^{(1)})$. This group is the symmetry group of this Painlev\'e equation, the formulas coincide with ones in \cite[Sec. 4]{Joishi:2015} after replacement
\[a_0=y_2^{-1}y_4^{-1}y_5^{-1},\;a_1=y_1^{-1}y_3^{-1},\;b=y_2^{-1}y_3^{-1}y_5^{-2},\;f_0=y_5,\;f_1=y_4^{-1},\quad w_0=\pi \sigma \pi, \;w_1=\sigma.\]
The invariant under the action of $\tilde{\mathcal{G}}_\mathcal{Q}$ is

In the autonomous case, $a_0,a_1,b$ are Casimirs functions, such that  $a_0a_1=1$. Variables $f_0,f_1$ are coordinates on the phase space with bracket $\{f_0,f_1\}=f_0f_1$. The invariant under the group $\tilde{\mathcal{G}}_\mathcal{Q}$ Hamiltonian has the form
%
\be
H=\sqrt{a_0 b}\, f_0+\sqrt{a_1b}\, f_1+\frac{\sqrt{a_0}}{\sqrt{b}\, f_0}+\frac{\sqrt{a_0}}{\sqrt{b}\, f_0f_1}+\frac{\sqrt{a_0}}{\sqrt{b}\, f_1}
\ee

Notice that, as usual, it is a Laurent polynomial of generally fractional powers of the dynamical variables, and
this is a general phenomenon -- see also other cases below. 

\paragraph{$\mathbf{A_5^{(1)}}$.} The group $\tilde{\mathcal{G}}_\mathcal{Q}$ contains elements
\be
s_1=(1,4)\circ \mu_4\circ\mu_1,\; s_2=(2,5)\circ \mu_5\circ\mu_2,\; s_0=(3,6)\circ \mu_6\circ\mu_3,
\; \pi=(1,2,3,4,5,6), \\ r_1=(3,5)\circ \mu_1\circ \mu_3\circ \mu_5 \circ \mu_1, \; r_0=(4,6)\circ \mu_2\circ \mu_4\circ \mu_6 \circ \mu_2,\;  \sigma=(1,4)(2,3)(5,6)\circ \varsigma.
\ee
The transformation $\pi$ acts as permutation of $y_1,\ldots,y_6$, the transformation $\sigma$ is a composition of inversion and permutation. The other elements have the form
\be
s_1\colon (y_1,y_2,y_3,y_4,y_5,y_6) \mapsto \left(y_4^{-1},y_2\frac{1+y_4}{1+y_1^{-1}},y_3\frac{1+y_4}{1+y_1^{-1}},y_1^{-1},
y_5\frac{1+y_1}{1+y_4^{-1}},y_6\frac{1+y_1}{1+y_4^{-1}}\right)
\\
s_2\colon (y_1,y_2,y_3,y_4,y_5,y_6) \mapsto \left(y_1\frac{1+y_2}{1+y_5^{-1}},y_5^{-1},y_3\frac{1+y_5}{1+y_2^{-1}},y_4\frac{1+y_5}{1+y_2^{-1}},
y_2^{-1},y_6\frac{1+y_2}{1+y_5^{-1}}\right)
\\
s_0\colon (y_1,y_2,y_3,y_4,y_5,y_6) \mapsto \left(y_1\frac{1+y_3}{1+y_6^{-1}},y_2\frac{1+y_3}{1+y_6^{-1}},y_6^{-1},
y_4\frac{1+y_6}{1+y_3^{-1}},y_5\frac{1+y_6}{1+y_3^{-1}},y_3^{-1}\right)
\\
r_1\colon (y_1,y_2,y_3,y_4,y_5,y_6) \mapsto \left(\frac{y_1}{y_3y_5}\frac{1{+}y_3{+}y_1^{-1}}{1{+}y_1{+}y_5^{-1}},
y_1y_2y_3\frac{1{+}y_5{+}y_3^{-1}}{1{+}y_1{+}y_5^{-1}},
\frac{y_3}{y_5y_1}\frac{1{+}y_5{+}y_3^{-1}}{1{+}y_3{+}y_1^{-1}},\right.\\ \quad\quad\quad\quad\quad\quad \left.y_3y_4y_5\frac{1{+}y_1{+}y_5^{-1}}{1{+}y_3{+}y_1^{-1}},
\frac{y_5}{y_1y_3}\frac{1{+}y_1{+}y_5^{-1}}{1{+}y_5{+}y_3^{-1}},
y_5y_6y_1\frac{1{+}y_3{+}y_1^{-1}}{1{+}y_5{+}y_3^{-1}}\right)
\\
r_0\colon (y_1,y_2,y_3,y_4,y_5,y_6) \mapsto \left(y_6y_1y_2\frac{1{+}y_4{+}y_2^{-1}}{1{+}y_6{+}y_4^{-1}},
\frac{y_2}{y_4y_6}\frac{1{+}y_4{+}y_2^{-1}}{1{+}y_2{+}y_6^{-1}},
y_2y_3y_4\frac{1{+}y_6{+}y_4^{-1}}{1{+}y_2{+}y_6^{-1}},\right.\\ \quad\quad\quad\quad\quad\quad \left.\frac{y_4}{y_2y_6}\frac{1{+}y_6{+}y_4^{-1}}{1{+}y_4{+}y_2^{-1}},
y_4y_5y_6\frac{1{+}y_2{+}y_6^{-1}}{1{+}y_4{+}y_2^{-1}},
\frac{y_6}{y_2y_4}\frac{1{+}y_2{+}y_6^{-1}}{1{+}y_6{+}y_4^{-1}}\right)
\ee
The elements $s_0,s_1,s_2$ generate affine Weyl group $W(A_2^{(1)})$, elements $r_0,r_1$ generate affine Weyl group $W(A_1^{(1)})$, these two Weyl groups commute. The conjugation by $\pi,\sigma$ preserves these subgroups, the full group generated by $\pi,\sigma,s_0,s_1,s_2,r_0,r_1$ is an extended affine Weyl group $\tilde{W}((A_2+A_1)^{(1)})$. This group is the symmetry group of this Painlev\'e equation, the formulas coincide with ones in \cite{Tsuda:2006:1} after replacement
\be
a_1=(y_1y_4)^{-1/2},\; a_2=(y_2y_5)^{-1/2},\; a_3=(y_3y_6)^{-1/2},\; b_1=(y_1y_3y_5)^{-1/2},\;b_0=(y_2y_4y_6)^{-1/2},\\ f=\left(\frac{y_6y_1^2y_2}{y_3y_4^2y_5}\right)^{1/6}=b_0^{-1/3}b_1^{1/3}a_1y_1,\ \ \ \ \ g=\left(\frac{y_1y_2^2y_3}{y_4y_5^2y_6}\right)^{1/6}=b_0^{1/3}b_1^{-1/3}a_2y_2.
\ee

In the autonomous case, $a_1,a_2,a_3,b_1,b_0$ are Casimirs functions, such that  $a_0a_1a_2=b_0b_1=1$. Variables $f,g$ are coordinates on the phase space with bracket $\{f,g\}=fg$. The invariant under the group $\tilde{\mathcal{G}}_\mathcal{Q}$ Hamiltonian has the form
\be
H=\sqrt[3]{a_1a_3^{2}b_1}\,f+\sqrt[3]{a_1^{2}a_2b_0}\,g+\frac{\sqrt[3]{a_1a_3^{2}b_0}}{f}+\frac{\sqrt[3]{a_1^{2}a_2b_1}}{g}+
\frac{\sqrt[3]{a_2^{2}a_3b_0}\,f}{g}+\frac{\sqrt[3]{a_2^{2}a_3b_1}\,g}{f}
\ee

\paragraph{$\mathbf{A_4^{(1)}}$.} The group $\tilde{\mathcal{G}}_\mathcal{Q}$ contains elements
\be
s_1=(1,2),\; s_2=(1,5)\circ \mu_1\circ\mu_5,\; s_3=(3,7)\circ \mu_3\circ\mu_7,\; s_4=(3,4),\\
s_0=(1,3)\circ\mu_6\circ\mu_3\circ\mu_1\circ\mu_6,\;\pi=(2,4,5,6,7)(1,3)\circ\mu_3,\; \sigma=(1,3)(2,4)(5,7)\circ\varsigma.
\ee
The transformations $s_1, s_3$ are just permutations, $\sigma$ is a composition of permutation and inversion, the other transformations look like
\be
s_0\colon (y_1,y_2,y_3,y_4,y_5,y_6,y_7)\mapsto \left(\frac{y_1}{y_3y_6}\frac{1{+}y_3{+}y_1^{-1}}{1{+}y_1{+}y_6^{-1}},
y_1y_2\frac{1{+}y_3{+}y_1^{-1}}{1{+}y_1{+}y_6^{-1}},
\frac{y_3}{y_1y_6}\frac{1{+}y_6{+}y_3^{-1}}{1{+}y_3{+}y_1^{-1}},\right. \\
\left.y_3y_4\frac{1{+}y_6{+}y_3^{-1}}{1{+}y_3{+}y_1^{-1}},y_3y_5y_6\frac{1{+}y_1{+}y_6^{-1}}{1{+}y_3{+}y_1^{-1}},\frac{y_6}{y_1y_3}\frac{1{+}y_1{+}y_6^{-1}}{1{+}y_6{+}y_3^{-1}},y_1y_6y_7\frac{1{+}y_3{+}y_1^{-1}}{1{+}y_6{+}y_3^{-1}} \right),
\\
s_2\colon (y_1,y_2,y_3,y_4,y_5,y_6,y_7)\mapsto \left(y_5^{-1},y_2,y_3\frac{1+y_5}{1+y_1^{-1}},y_4\frac{1+y_5}{1+y_1^{-1}},
y_1^{-1},y_6\frac{1+y_1}{1+y_5^{-1}},y_7\frac{1+y_1}{1+y_5^{-1}}\right),
\\
s_3\colon (y_1,y_2,y_3,y_4,y_5,y_6,y_7)\mapsto \left(y_1\frac{1+y_3}{1+y_7^{-1}},y_2\frac{1+y_3}{1+y_7^{-1}},y_7^{-1},
y_4,y_5\frac{1+y_7}{1+y_3^{-1}},y_6\frac{1+y_7}{1+y_3^{-1}},y_3^{-1}\right),
\\
\pi\colon (y_1,y_2,y_3,y_4,y_5,y_6,y_7)\mapsto \left(y_3^{-1},y_7,y_1(1+y_3),y_2(1+y_3),y_4,\frac{y_5}{1+y_3^{-1}},\frac{y_6}{1+y_3^{-1}}\right).
\ee
The transformations $s_0,\ldots,s_4$ generate group isomorphic to Weyl group of the affine roots system $E_4^{(1)}=D_4^{(1)}$. And if we add $\pi$ we get an extended Weyl group. This group is the symmetry group of this Painlev\'e equation, the formulas coincide with ones in \cite[App. A]{Tsuda:2006:1} after replacement
\be
	a_0=y_1y_3y_6,\;a_1=y_1^{-1}y_2,\; a_2=y_1y_5,\; a_3=y_3y_7,\; a_4=y_3^{-1}y_4,\\ f=\frac{y_3y_6}{y_5y_7}=\frac{a_0}{a_2}y_7^{-1},\;g=\frac{y_7}{y_6}=\frac{a_3}{a_0}y_1.
\ee

In the autonomous case, $a_0,a_1,a_2,a_3,a_4$ are Casimirs functions, such that  $a_0a_1a_2a_3a_4=1$. Variables $f,g$ are coordinates on the phase space with bracket $\{g,f\}=gf$. The invariant under the group~$\tilde{\mathcal{G}}_\mathcal{Q}$ Hamiltonian has the form

\be
H=\sqrt[5]{\frac{a_0^{12}a_1^{4}}{a_2^{4}a_3^{7}}}\frac gf+\sqrt[5]{\frac{a_0^{2}a_3^{3}}{a_1a_2^{4}}}\frac1{fg}+
\left(\sqrt[5]{\frac{a_0^{2}}{a_1a_2^{4}a_3^{7}}}+\sqrt[5]{\frac{a_0^{7}a_1^{4}a_2}{a_3^{2}}}\right)g+\\+
\left(\sqrt[5]{\frac{a_0^{7}}{a_1a_2^{4}a_3^{2}}}+
\sqrt[5]{\frac{a_0^{7}a_1^{4}}{a_2^{4}a_3^{2}}}\right)\frac1f+
\sqrt[5]{\frac{a_2a_3^{3}}{a_0^{3}a_1}}\frac1g+\sqrt[5]{\frac{a_2}{a_0^{3}a_1a_3^{2}}}\, fg+\sqrt[5]{\frac{a_2^{6}a_3^{3}}{a_0^{8}a_1}}\, f
\ee

\paragraph{$\mathbf{A_3^{(1)}}$.} The group $\tilde{\mathcal{G}}_\mathcal{Q}$ contains elements
\be
s_0=(1,2),\;\; s_1=(5,6),\;\; s_2=(1,5)\circ\mu_5\circ\mu_1,\;\; s_3=(3,7)\circ \mu_3\circ\mu_7,\\ s_4=(3,4),\;\; s_5=(7,8),\;\;
\pi=(1,7,5,3)(2,8,6,4),\; \sigma=(1,7)(2,8)(3,5)(4,6)\circ \varsigma.
\ee
The transformations $s_0, s_1, s_4, s_5, \pi$ are permutations of $y_1,\ldots, y_8$, $\sigma$ is a composition of inversion and permutations. The  only tricky elements are the reflections $s_2,s_3 $,	
\be
s_2\colon (y_1, y_2, y_3, y_4,y_5,y_6,y_7,y_8)\mapsto \left(y_5^{-1},y_2,y_3\frac{1+y_5}{1+y_1^{-1}},y_4\frac{1+y_5}{1+y_1^{-1}},y_1^{-1},y_6,
y_7\frac{1+y_1}{1+y_5^{-1}},y_8\frac{1+y_1}{1+y_5^{-1}}\right),\\
s_3\colon (y_1, y_2, y_3, y_4,y_5,y_6,y_7,y_8)\mapsto \left(y_1\frac{1+y_3}{1+y_7^{-1}},y_2\frac{1+y_3}{1+y_7^{-1}},y_7^{-1},y_4,y_5\frac{1+y_7}{1+y_3^{-1}},
y_6\frac{1+y_7}{1+y_3^{-1}},y_3^{-1},y_8\right).
\ee
The elements $s_0,s_1,\ldots,s_5$ generate the group, isomorphic to the Weyl group of the affine roots system $E_5^{(1)}=D_5^{(1)}$, and adding $\pi$ one gets an extended Weyl group. This group is the symmetry group of this Painlev\'e equation, the formulas coincide with ones in \cite[Sec. 1]{TsudaMasuda} after replacement
\be
a_0=\sqrt[4]{\frac{y_2}{y_1}},\; a_1=\sqrt[4]{\frac{y_6}{y_5}},\; a_2=\sqrt[4]{y_1y_5},\;a_3=\sqrt[4]{y_3y_7},\;a_4=\sqrt[4]{\frac{y_4}{y_3}},\; a_5=\sqrt[4]{\frac{y_8}{y_7}},
\\
\;f=\sqrt[4]{\frac{y_7y_8}{y_3y_4}}=\frac{a_5}{a_3^2a_4}y_7,\quad g=\sqrt[4]{\frac{y_5y_6}{y_1y_2}}=\frac{a_1}{a_0a_2^2}y_5,
\ee
and $\sigma=\sigma_1$, $\pi=\sigma_1\circ \sigma_2$.

In the autonomous case, $a_0,a_1,a_2,a_3,a_4,a_5$ are Casimirs functions, such that  $a_0a_1a_2^2a_3^2a_4a_5=1$. Variables $f,g$ are coordinates on the phase space with bracket $\{g,f\}=fg$. The invariant under the group $\tilde{\mathcal{G}}_\mathcal{Q}$ Hamiltonian has the form
\be
H=\left(a_0^{2}+\frac1{a_0^2}\right)f+\left(a_5^{2}+\frac1{a_5^2}\right)g+\left(a_1^{2}+\frac1{a_1^2}\right)\frac1{f}+
\left(a_4^{2}+\frac1{a_4^2}\right)\frac1g+\\+\frac{1}{a_0a_1a_2^{2}}\left(fg+\frac1{fg}\right)+a_0a_1a_2^{2}\left(\frac fg+\frac gf\right)
\ee

\paragraph{$\mathbf{A_2^{(1)}}$.} The group $\tilde{\mathcal{G}}_\mathcal{Q}$ contains elements
\be
s_1=(2,3),\;\; s_2=(1,2),\;\; s_4=(4,5),\;\; s_5=(5,6),\;\; s_6=(7,8),\;\; s_0=(8,9),\\
s_3=(4,7)\circ\mu_1\circ\mu_4\circ\mu_7\circ\mu_1,\;\; \pi=(1,4,7)(2,5,8)(3,6,9),\; \sigma=(1,7)(2,8)(3,9)\circ \varsigma.
\ee
The transformations $s_0, s_1, s_4, s_5, \pi$ are permutations of $y_1,\ldots, y_8$, $\sigma$ is a composition of inversion and permutations. The  only tricky element is the reflections $s_3$	
\begin{multline}
	s_3\colon (y_1, y_2, y_3, y_4,y_5,y_6,y_7,y_8,y_9)\mapsto \left(\frac{y_1}{y_4y_7}\frac{1{+}y_4{+}y_1^{-1}}{1{+}y_1{+}y_7^{-1}},y_1y_2\frac{1{+}y_4{+}y_1^{-1}}{1{+}y_1{+}y_7^{-1}},y_1y_3\frac{1{+}y_4{+}y_1^{-1}}{1{+}y_1{+}y_7^{-1}}, \frac{y_4}{y_1y_7}\frac{1{+}y_7{+}y_4^{-1}}{1{+}y_4{+}y_1^{-1}},
	\right. \\
	\left. y_4y_5\frac{1{+}y_7{+}y_4^{-1}}{1{+}y_4{+}y_1^{-1}},y_4y_6\frac{1{+}y_7{+}y_4^{-1}}{1{+}y_4{+}y_1^{-1}},
	\frac{y_7}{y_1y_4}\frac{1{+}y_1{+}y_7^{-1}}{1{+}y_7{+}y_4^{-1}},
	y_7y_8\frac{1{+}y_1{+}y_7^{-1}}{1{+}y_7{+}y_4^{-1}},
	y_7y_9\frac{1{+}y_1{+}y_7^{-1}}{1{+}y_7{+}y_4^{-1}}\right).
\end{multline}
The elements $s_0,s_1,\ldots,s_6$ generate subgroup isomorphic to Weyl group of the affine roots system $E_6^{(1)}$. And if we add $\pi,\sigma $ we get an extended Weyl group. This group is the symmetry group of this Painlev\'e equation, the formulas coincide with ones in \cite[Sec. 3]{Tsuda:2008} after replacement
\be
a_1=\sqrt[3]{\frac{y_2}{y_3}},\;a_2=\sqrt[3]{\frac{y_1}{y_2}},\; a_3=\sqrt[3]{\frac1{y_1y_4y_7}},\; a_4=\sqrt[3]{\frac{y_4}{y_5}},\;a_5=\sqrt[3]{\frac{y_5}{y_6}},\; a_6=\sqrt[3]{\frac{y_7}{y_8}},\;a_0=\sqrt[3]{\frac{y_8}{y_9}},\\
 f=\sqrt[9]{\frac{y_1^2y_2^2y_3^2}{y_4y_5y_6y_7y_8y_9}}=a_1^{-1}a_2^{-2}q^{-1/9}y_1,\quad g=\sqrt[9]{\frac{y_1y_2y_3y_4y_5y_6}{y_7^2y_8^2y_9^2}}=a_6^2a_0q^{1/9}y_7^{-1}, 
\ee
and $i_1= \pi\circ \varsigma$, $i_2=\varsigma$. Here $q=a_0^{-3}a_1^{-3}a_2^{-6}a_3^{-9}a_4^{-6}a_5^{-3}a_6^{-6}$, in accordance to \eqref{eq:q} (and slightly different from $q$ in \cite{Tsuda:2008}).

In the autonomous case, $a_0,a_1,a_2,a_3,a_4,a_5,a_6$ are Casimirs functions, such that  $a_0a_1a_2^{2}a_3^{3}a_4^{2}a_5a_6^{2}=1$. Variables $f,g$ are coordinates on the phase space with bracket $\{f,g\}=fg$. The invariant under the group $\tilde{\mathcal{G}}_\mathcal{Q}$ Hamiltonian has the form
\be
H=a_4^{2}a_5f+a_4a_5^{2}g+\frac{1}{a_0a_6^{2}f}+\frac{1}{a_1^{2}a_2g}+\frac{1}{fg}+\frac{a_0^{2}a_6}{f}+\frac{a_0a_6^{2}g}{f}+\frac{a_0g}{a_6f}
+\frac{a_1^{2}a_2f}{g}+\frac{a_1a_2^{2}}{g}+\\+\frac{a_1}{a_2g}+\frac{a_2f}{a_1g}+\frac{a_4g}{a_5}+\frac{a_5f}{a_4}+\frac{a_6}{a_0f}+\frac{f^{2}}{g}+
\frac{f}{a_1a_2^{2}g}+\frac{f}{a_4a_5^{2}}+\frac{g^{2}}{f}+\frac{g}{a_0^{2}a_6f}+\frac{g}{a_4^{2}a_5}
\ee

\subsection{$q$-Painlev\'e equations and cluster algebras} \label{ssec:qP:tau}

There is an alternative, or dual to the Poisson $X$-cluster varieties language, called
$A$-cluster varieties or just cluster algebras, see e.g. \cite{FZ:2006}. We call the corresponding $A$-cluster variables as $\{\tau_I\}$ due to their relation to the tau-functions for $q$-difference Painlev\'e equations.
Under mutation at $j$-th vertex these variables are transformed as
\be
\label{eq:muttau}
\mu_j:\ \  \tau_j\mapsto \frac{\prod_{b_{Ij}>0} \tau_I^{b_{Ij}}+\prod_{b_{Ij}<0} \tau_I^{-b_{Ij}}}{\tau_j}
\\
\tau_I\mapsto \tau_I,\ \ \ \ \ I\neq j
\ee
and antisymmetric matrix $B=\{b_{i,j}\}$ is transformed by the formula \eqref{mute}.

Compare to \rf{mutz}, there are several important distinctions. Generally there are more $\{\tau_I\}$-variables, than their $X$-cluster $\{y_i\}$-relatives. Mutations \rf{eq:muttau} are allowed only in the vertices of $\Gamma$, other vertices of the extended quiver $\hat{\Gamma}$ are called therefore frozen.
A relation between $\tau_I$ and $y_j$ is given by the formula
\begin{equation}\label{eq:ytau}
y_j=\prod_{I \in \hat{\Gamma}} \tau_I^{b_{Ij}},
\end{equation}
it is easy to verify that mutation of $y_j$ defined in such way coincides with \eqref{mutz}.

The elements $b_{IJ}$ of the exchange matrix $B$, with $I,J \in \hat{\Gamma}\setminus\Gamma$ do not enter in the formulas for mutation of $\tau_j$. So usually the matrix $B$ is assumed to be rectangular  $B=\{b_{Ij}\}$, $I=1,\ldots,|\hat{\Gamma}|$, $j=1,\ldots,|\Gamma|$, where its square part $\left.B\right|_\Gamma=\{\epsilon_{ij}\}$ is the  exchange matrix of the $X$-cluster Poisson variety.

	\begin{Remark}
	One can get rid of the frozen vertices by introducing into \rf{eq:muttau} ``coefficients'' $\{\tilde{y}_i\}$ (in terminology of \cite{FZ:2006}, where $\{\tau_i\}$ are called just ``cluster variables'')
	\be
	\label{eq:muttauy}
	\mu_j:\ \  \tau_j\mapsto \frac{\tilde{y}_j\prod_{\epsilon_{ij}>0} \tau_i^{\epsilon_{ij}}+\prod_{\epsilon_{ij}<0} \tau_i^{-\epsilon_{ij}}}{(1\oplus \tilde{y}_j)\tau_j},\ \ \ \ \  \tau_i\mapsto \tau_i,\ \ \  i\neq j
	\ee
	where the coefficients $\{\tilde{y}_i\}$ are direct analogs of the coordinates $\{y_i\}$ on $X$-cluster variety, but belong to certain semifield \cite{FZ:2006}, the sign $\oplus$ means addition in this semifield (one should also use it in the corresponding transformation of $\{\tilde{y}_i\}$, according to the formula \eqref{mutz}).
	There are two main cases of the semifield: universal semifield, in this case one can think that $\{\tilde{y}_i\}$ are just numbers, and tropical semifield, this case can be modeled using the frozen cluster variables $\{\tau_I|I\in \hat{\Gamma}\setminus\Gamma\}$.
	\end{Remark}

\paragraph{$\mathbf{A_7^{(1)'}}$.}  Consider an example of ``relativistic Toda'', here the exchange matrix $B=\{b_{Ij}\}$ of extended quiver $\hat{\Gamma}$  has the form
\be
B= \begin{pmatrix}
	0 & 2 & 0 & -2 \\
	-2 & 0 & 2 & 0 \\
	0 & -2 & 0 & 2 \\
	2 & 0 & -2 & 0 \\
	1 & 0 & 1 & 0 \\
	
	1 & -1 & 1 & -1 \\
	-1 & 0 & -1 & 0 \\
	-1 & 1 & -1 & 1
\end{pmatrix}
\ee
whose upper $4\times 4$ part $\left.B\right|_\Gamma=\{\epsilon_{ij}\}$ coincides with the exchange matrix of the quiver $A_7^{(1)'}$, see
Fig.~\ref{fi:PQ}. We associate four unfrozen variables $\{\tau_1, \tau_2, \tau_3, \tau_4\}$ with the vertices
of this quiver, and add four extra frozen $\{\tau_5, \tau_6, \tau_7, \tau_8\}$.

Let us also introduce separate notations for new frozen variables: $\tau_5=q_0$, $\tau_6=z_0$, $\tau_7=q_1$ and $\tau_8=z_1$. The operator $T=(1,2)(3,4)\mu_1\mu_3$ (already used above and defined in \rf{Trt1}, \rf{Trt2}) according to \rf{eq:muttau} acts on the corresponding $A$-cluster variables as

\begin{equation}
(\tau_1, \tau_2, \tau_3, \tau_4)\mapsto (\tau_2,\frac{q_0z_0\tau_4^2  + q_1z_1\tau_2^2}{\tau_1}, \tau_4, \frac{q_1z_1\tau_4^2 + q_0z_0\tau_2^2}{\tau_3})
\end{equation}
This dynamics is equivalent to $q$-difference Painlev\'e equation for the tau-functions with surface type $A_7^{(1)'}$ as written in \cite[Table 1]{BS:2016:1} after replacement $q_0=q^{1/4}$, $q_1=q^{-1/4}$, $z_0=Z^{1/4}$, $z_1=Z^{-1/4}$. One can also check that action of transformation $T$ on frozen part of the quiver is equivalent to multiplication $z \mapsto qz$.

It will be convenient to take another matrix
\be
\label{Bmatrix}
B= \begin{pmatrix}
	0 & 2 & 0 & -2 \\
	-2 & 0 & 2 & 0 \\
	0 & -2 & 0 & 2 \\
	2 & 0 & -2 & 0 \\
	2 & 0 & 2 & 0 \\
	2 & -2 & 2 & -2
\end{pmatrix}
\ee
Its upper $4\times 4$ part $\left.B\right|_\Gamma=\{\epsilon_{ij}\}$ is the same, but the ``frozen'' is a bit different. Denote frozen variables in this case by $\tau_5=q^{1/4}$, $\tau_6=Z^{1/4}$.  Denote the action of $T$ as overline, and action of $T^{-1}$ as underline. Then we have
\begin{equation}\label{eq:Ttau}
\overline{(\tau_1, \tau_2, \tau_3, \tau_4)}=(\tau_2,\frac{\tau_2^2  + q^{1/2}Z^{1/2}\tau_4^2}{\tau_1}, \tau_4, \frac{\tau_4^2 + q^{1/2}Z^{1/2}\tau_2^2}{\tau_3})
\end{equation}
These leads to bilinear equations
\begin{equation}\label{eq:bilintau}
\underline{\tau_1}\overline{\tau}_1=\tau_1^2+Z^{1/2}\tau_3^2,\quad  \underline{\tau_3}\overline{\tau}_3=\tau_3^2+Z^{1/2}\tau_1^2.
\end{equation}
One can consider the $\overline{\tau_i}=\tau_i(qZ)$, $\underline{\tau_i}=\tau_i(q^{-1}Z)$, then the equations \eqref{eq:bilintau} become $q$-difference bilinear equations. These equations can be called the bilinear form of the $q$-Painlev\'e equation (of the surface type $A_7^{(1)'}$). Namely, if we define $G=Z^{1/2}\tau_3^2\tau_1^{-2}$ following \eqref{eq:GA7p} and \eqref{eq:ytau} we get that $G$ satisfies \eqref{eq:qPAtp}.

In the paper \cite{BS:2016:1} the formal solution of these equations was proposed, namely $\tau_1=\mathcal{T}(u,s;q|Z)$,
$\tau_3=is^{1/2}\mathcal{T}(uq,s;q|Z),
$ where
\begin{equation}\label{eq:T}
\mathcal{T}(u,s;q|Z)=\sum_{m \in \mathbb{Z}}s^m \mathsf F(uq^{2m};q,q^{-1}|Z).
\end{equation}
Here $\mathsf F(u;q,q^{-1};Z)$ is a properly normalized $q$-deformed conformal block, see Appendix \ref{App:q-functions} for the definition, $u,s$ parameterize solutions of second order $q$-difference equation \eqref{eq:qPAtp} (integration constants), see also Remark~\ref{rem:us:bracket} below.

The formula \eqref{eq:T} is a $q$-analogue of the proposed in \cite{GIL1207},\cite{GIL1302} ``Kiev-formulas'' for the tau-function of differential Painlev\'e equations in terms of $c=1$ conformal blocks.

\begin{Remark} \label{rem:us:bracket}
	It is natural to ask what is meaning of the cluster Poisson bracket for the parameters $u,s$. This question is also important for the quantization which we will study below. It turns out that $u$ and $s$ are log-canonically conjugated coordinates (similarly to the differential case)
	
	Cluster Poisson bracket can be written as $\{G,F\}=2GF$. Substituting
	$G=G(Z)$ and $F=G(q^{-1}Z)^{-1}$ we come to the non-trivial functional relation $\{\log G(q^{-1}Z),\log G(Z)\}=f(Z)=2$. We check that it satisfied
	if we use \eqref{eq:T}, \eqref{eq:ytau}, and put
	\begin{equation}
		\{s,u\}=2su
	\end{equation}
	The proof of this relation goes like follows: first we notice that $f(Z)=f(qZ)$ since $G(Z)$ satisfies \eqref{eq:qPAtp}, and then we check by
	explicit calculation that $f(Z)$ is a power series, therefore $f(Z)$ is constant. Then compute that $f(Z)=f(0)=2$.

\end{Remark}

\paragraph{Other $q$-Painlev\'e equations.} For the $q$-Painlev\'e equations $\mathbf{A_7^{(1)}}$, $\mathbf{A_6^{(1)}}$, $\mathbf{A_5^{(1)}}$, $\mathbf{A_4^{(1)}}$, $\mathbf{A_3^{(1)}}$, $\mathbf{A_2^{(1)}}$ we believe that similar construction with frozen variables can reproduce dynamics of tau-functions. This is easier to check in the coefficient free case, where all parameters of $q$-Painlev\'e equation are set to be unit. In terms of the cluster algebra it means, that we have no nontrivial frozen variables, just the quiver from the Fig.~\ref{fi:PQ}. One can also say, that $\{\tilde{y}_i=1\}$ in tropical semifield, or this is the integrable $q=1$ case, moreover restricted to a special case, when all Casimir functions are put to unities. In this case action of the subgroup of $\mathcal{G}_\mathcal{Q}$ presented above is equivalent to the action on tau-functions, see \cite{Joishi:2015} for the case of $A_6^{(1)}$ surface, \cite{Tsuda:2006:1} for the case of $A_4^{(1)}$ and $A_5^{(1)}$ surfaces, \cite{TsudaMasuda} for the case $A_3^{(1)}$ surface, \cite{Tsuda:2006:2} for the case $A_2^{(1)}$ surface.

The ``Kiev formula" for the $A_3^{(1)}$ surface was proposed in \cite{Jimbo:2017}, see also \cite{Morozov:2017} and \cite{BGT2} (see Discussion).

\section{Quantization
	\label{ss:quant}}

\subsection{Quantization of the $X$-cluster variety}

Consider now quantization of the $X$-cluster varieties, following \cite{FG:2003}. For the logarithmically constant Poisson bracket \rf{PB} this is an obvious canonical procedure of quantum mechanics, and we will not even distinguish notations for the coordinates on $X$-cluster variety and their quantum analogs.
We denote the multiplicative quantization parameter as $p$ in order to distinguish it from the parameter $q$ in difference equations. We do not impose any relation on $p,q$, at the end of this section it will be convenient to express them $p=q_1^2q_2^2, q=q_2^2$ in terms of Nekrasov background parameters $q_1, q_2$ (see also Appendix~\ref{App:q-functions} for notations).

The quantization of the quadratic Poisson bracket \eqref{PB} has the form
\be
\label{eq:yy=pyy}
y_i y_j=p^{-2\epsilon_{ij}}y_jy_i
\ee
of the quadratic relations for the generators of the quantum algebra of functions $Fun_p(\mathcal{X})$.
Quantum mutations $\mu_j$ for these generators are given by (compare to \eqref{mutz})
\be
\mu_j:\ \ 	y_j\mapsto y_j^{-1},\quad y_i^{1/|\epsilon_{ij}|} \mapsto y_i^{1/|\epsilon_{ij}|}\left(1+p y_j^{{\rm sgn\,} \epsilon_{ij}}\right)^{{\rm sgn\,} \epsilon_{ij}},\ i\neq j
\ee
while the mutation of the quiver itself is still given by formula \eqref{mute} for the exchange matrix $\epsilon$. One can check that  mutations of $\{y_i\}$ and $\epsilon$ preserve the relations \eqref{eq:yy=pyy}.
\begin{Remark}
	These formulas are actually equivalent to those of \cite[Lemma 3.5]{FG:2003}:
	\begin{equation}
y_i \mapsto \left\{\begin{aligned}
&y_j^{-1},&\quad &\text{for } i=j;\\
&y_i(1+py_j) (1+p^3y_j)\cdot \ldots \cdot (1+p^{2\epsilon_{ij}-1}y_j),& \quad &\text{for } \epsilon_{ij}\geq 0\\
&y_i\left((1+py_j^{-1}) (1+p^3y_j^{-1})\cdot \ldots \cdot (1+p^{2|\epsilon_{ij}|-1}y_j^{-1})\right)^{-1}, &\quad &\text{for } \epsilon_{ij}< 0
\end{aligned}
\right.
	\end{equation}
up to replacement $\epsilon_{ij} \leftrightarrow -\epsilon_{ij}$, but have more compact form.
\end{Remark}
Below in this section we restrict ourselves again to the case of the quiver $A_7^{(1)'}$ from the Fig.~\ref{fi:PQ}. As was already mentioned many times (see Sect.~\ref{ssec:2.3}) in the autonomous case $q=1$ this quiver corresponds to classical relativistic Toda chain with two particles.

\subsection{Quantum Toda and quantum Painlev\'e}
Notice first, that there are two central or Casimir elements
\be
Z=y_1y_3,\ \ \ \ \
q=y_1y_3y_2y_4
\ee
in the quantum function algebra $Fun_p(\mathcal{X})$, with relations \eqref{eq:yy=pyy} with the exchange matrix of the quiver $A_7^{(1)'}$.
Similarly to classical case consider the discrete flow $ T=(1,2)(3,4)\circ \mu_1\circ \mu_3$. In quantum case it can be defined as a map

\begin{align}
\label{Tp}
\left(y_1^{1/2},y_2^{1/2},y_3^{1/2},y_4^{1/2}\right)\mapsto \left(y_2^{1/2}\frac{1+py_3}{1+py_1^{-1}},\,y_1^{-1/2},\,y_4^{1/2}\frac{1+py_1}{1+py_3^{-1}},\,y_3^{-1/2}\right).
\end{align}
where the ratios in the r.h.s. are well-defined, since $y_1$ and $y_3$ commute with each other due to \eqref{eq:yy=pyy} (and certainly commute with the quantization parameter $p$). This action on Casimirs has the form
\be
\label{TpqZ}
T\colon (q,Z) \mapsto (q, qZ)
\ee

For $q=1$ we have an invariant Hamiltonian
\begin{equation}
	H=
	y_2^{1/2}y_1^{1/2}+y_1^{1/2}y_2^{-1/2}+y_2^{-1/2}y_1^{-1/2}+Zy_1^{-1/2}y_2^{1/2}
\end{equation}
This is a Hamiltonian of quantum relativistic two-particle affine Toda chain.\footnote{Consider the space functions on $x_1, x_2$ depending only on the difference $x_1-x_2$. Let $T_1,T_2$ the shifts of $x_1,x_2$ correspondingly, on this space of functions $T_1=T_2^{-1}=T$ and $Te^{x_1-x_2}=pe^{x_1-x_2}T$. Then
	\begin{equation}
	p^{-1/2}H=T^2+T^{-2}+e^{x_1-x_2}+Ze^{x_2-x_1}
	\end{equation}
	where \[T^2=p^{-1/2}y_2^{1/2}y_1^{1/2}, \quad e^{x_1-x_2}=p^{-1/2}y_1^{1/2}y_2^{-1/2}.\]
	The last  formula for $H$ is equivalent  \cite[eq. (6.3)]{Etingof:1999}}

Now let $q\neq 1$, in other words consider deautonomization of quantum Toda. Denote $G=y_1$, $F=y_2$, then due to  \eqref{eq:yy=pyy} one gets  $FG=p^{4}GF$. Denote also the action \rf{Tp} of $T$ as overline, and action of $T^{-1}$ as underline.
Then from \rf{Tp} one gets
\begin{equation}
	\overline{G}^{1/2}=F^{1/2}\frac{G+pZ}{G+p},\quad
	\overline{F}=G^{-1}
\end{equation}
therefore
\be\label{eq:quantumP3G12}
\left\{\begin{aligned}\underline{G}^{1/2} \;	\overline{G}^{1/2}&=\frac{G+pZ}{G+p}, \\
	G\underline{G}&=p^4\underline{G}G.
\end{aligned}\right.
\ee
It also follows from these equation, that
\begin{equation}\label{eq:quantumP3}
\underline{G}	\overline{G}=\frac{(G+pZ)(G+p^3Z)}{(G+p)(G+p^3)}
\end{equation}
We propose to call this relation as \emph{quantum} $q$-Painlev\'e equation, its classical analogue is \eqref{eq:qPAtp}.

\subsection{Quantum Painlev\'e and bilinear relations
	\label{ssec:quantF}}

Now we are almost ready to formulate the main result of this section. We claim that there is a formal solution
of the quantum $q$-Painlev\'e equation, i.e. of the equations \eqref{eq:quantumP3G12} and \eqref{eq:quantumP3}.
Our main result consists of two following statements:
\begin{itemize}
	\item There are quantum analogs of the $A$-cluster $\tau$-variables, which have been discussed in
	Sect.~\ref{ssec:qP:tau}, and we call them $\mathcal{T}$-functions of quantum $q$-Painlev\'e.
	Quantization of $A$-cluster algebra was proposed in \cite{BerZel:2004}, here we actually follow their approach. We define the action of discrete flow on the quantum $\mathcal{T}$-functions and show, that they satisfy bilinear relations, leading to \eqref{eq:quantumP3G12} and \eqref{eq:quantumP3}.
	
	\item These bilinear equations for the quantum $q$-difference Painlev\'e $\mathcal{T}$-functions by quantum analog of the formula \eqref{eq:T} turn into bilinear relations for Nekrasov functions in arbitrary $\Omega$-background, or for conformal blocks with arbitrary central charge. Hence, it is proposed quantization of the $q$-Painlev\'e system, which leads to the analog of the "Kiev formula" for generic non-integer
	central charge.
\end{itemize}

Hence, let us first turn to the tau-form of the quantum $q$-difference Painlev\'e equation. Following
\cite{BerZel:2004} we quantize $\{\tau_I\}$-variables and consider them as elements of the quantum cluster algebra with relations
\be
\label{tauLam}
\tau_I\tau_J=p^{\Lambda_{IJ}/2}\tau_J\tau_I,
\ee
where $I,J=1,\ldots,6$ and the matrix $\Lambda$ is \be
\label{Lam}
\Lambda=\begin{pmatrix}
	0 & 0 & 0 & 1 & 1 & 0\\
	0 & 0 & -1 & 0 & 1 & -1\\
	0 & 1 & 0 & 0 & 1 & 0\\
	-1 & 0 & 0 & 0 & 1 & -1\\
	-1 & -1 & -1 & -1 & 0 & 0\\
	0 & 1 & 0 & 1 & 0 & 0
\end{pmatrix}
\ee
For matrix $B$ defined in \eqref{Bmatrix} we have $\sum b_{iJ}\Lambda_{JI}=-4\delta_{iI}$, in other words $(-\Lambda,B)$ form a compatible pair in sense of \cite{BerZel:2004}.

For quantum $\{\tau_I\}$-variables we now fix the notations $\tau_1=\mathcal{T}_1$, $\tau_2=\mathcal{T}_2$, $\tau_3=\mathcal{T}_3$, $\tau_4=\mathcal{T}_4$, so that first four will be  quantum $\mathcal{T}$-functions. Two last are $\tau_5=q^{1/4}$, $\tau_6=Z^{1/4}$, they are still generally noncommutative with $\mathcal{T}_i$, as follows from \rf{tauLam}). We now define the discrete dynamics of the quantum $\mathcal{T}$-functions by
\be
\label{taumutq}
\overline{(\mathcal{T}_1,\mathcal{T}_2,\mathcal{T}_3,\mathcal{T}_4,Z,q)}= (\mathcal{T}_2,\mathcal{T}_1^{-1}(\mathcal{T}_2^2+p^2(qZ)^{1/2}\mathcal{T}_4^2),\mathcal{T}_4,
\mathcal{T}_3^{-1}(\mathcal{T}_4^2+p^2(qZ)^{1/2}\mathcal{T}_2^2),Zq,q),
\\
\underline{(\mathcal{T}_1,\mathcal{T}_2,\mathcal{T}_3,\mathcal{T}_4,Z,q)}= ((\mathcal{T}_1^2+p^2Z^{1/2}\mathcal{T}_3^2)\mathcal{T}_2^{-1},\mathcal{T}_1,(\mathcal{T}_3^2+
p^2Z^{1/2}\mathcal{T}_1^2)\mathcal{T}_4^{-1},\mathcal{T}_3,Zq^{-1},q).
\ee
It is straightforward to check that this dynamics preserves commutation relations \eqref{tauLam}.

The formulas \eqref{taumutq} lead to bilinear relations

\be
\label{qPb}
\underline{\mathcal{T}_1}\overline{\mathcal{T}_1}=\mathcal{T}_1^2+p^2Z^{1/2}\mathcal{T}_3^2,
\quad\underline{\mathcal{T}_3}\overline{\mathcal{T}_3}=\mathcal{T}_3^2+p^2Z^{1/2}\mathcal{T}_1^2,
\ee
These relation are quantum analogs of \eqref{eq:bilintau}

\begin{prop} \label{th:quantP}
	Let $G=pZ^{1/2}\mathcal{T}_1^2\mathcal{T}_3^{-2}$. Then, the bilinear equations \rf{qPb} for the quantum $\mathcal{T}$-functions imply the quantum $q$-difference Painlev\'e equations \eqref{eq:quantumP3G12} and \eqref{eq:quantumP3}.
\end{prop}

Note that elements $Z, \mathcal{T}_1, \mathcal{T}_3$ commute, so the order of factors in the definition of $G$ can be arbitrary.

The proof is straightforward. Due to \rf{tauLam} the element $G$ commutes with $Z,q$.
We have
\be
\underline{G}=pq^{-1/2}Z^{1/2}\bigl((\mathcal{T}_1^2+p^2Z^{1/2}\mathcal{T}_3^2)\mathcal{T}_2^{-1}\bigr)^2\bigl((\mathcal{T}_3^2+p^2Z^{1/2}\mathcal{T}_1^2)\mathcal{T}_4^{-1}\bigr)^{-2},\quad \overline{G}=pq^{1/2}Z^{1/2}\mathcal{T}_2^2\mathcal{T}_4^{-2}
\ee
From the second formula we get the commutation relation $\underline{G}G=p^4G\underline{G}$
which is equivalent to the second formula of \eqref{eq:quantumP3G12}.

Since the factors in $G$ commute one calculate square root and get
\be
\underline{G}^{1/2}=p^{1/2}q^{-1/4}Z^{1/4}\bigl((\mathcal{T}_1^2+p^2Z^{1/2}\mathcal{T}_3^2)
\mathcal{T}_2^{-1}\bigr)\bigl((\mathcal{T}_3^2+p^2Z^{1/2}\mathcal{T}_1^2)\mathcal{T}_4^{-1}\bigr)^{-1},
\\
\overline{G}^{1/2}=p^{1/2}q^{1/4}Z^{1/4}\mathcal{T}_2\mathcal{T}_4^{-1}.
\ee
Therefore
\begin{align}
\underline{G}^{1/2}\overline{G}^{1/2}=p^2Z^{1/2}(\mathcal{T}_1^2+p^2Z^{1/2}\mathcal{T}_3^2)(p^2\mathcal{T}_3^2+p^2Z^{1/2}\mathcal{T}_1^2)^{-1}=(G+p^{3}Z)(G+p)^{-1}
\end{align}
So we get the first formula of \eqref{eq:quantumP3G12} up to replacement $Z\mapsto p^{-2}Z$.

Now we want to present explicit formula for the $\mathcal{T}_i$. Denote $q_2=q^{1/2}$, $q_1=q_2^{-1}p^{2}$. The solution will be the function depending on variables $q_1,q_2,u,s,Z,a,b$ where the full set of nontrivial commutation relations for these variables is
\be
\label{eq:quantumparameters}
q_2^2a=p^{-2}aq_2^2=aq_1^{-1}q_2,\quad q_1q_2^{-1}a=p^2aq_1q_2^{-1}=aq_1^{2},\quad u s=p^4 s u,\quad Zb=p^2bZ.
\ee
The variables $u,s$ are quantum analogs of the $u,s$ used in Section \ref{ssec:qP:tau}, their commutation relation is the quantization of the Poisson bracket, see Remark \ref{rem:us:bracket}. The variables $a,b$ not commuting with $q_2,Z$ are necessary, since $\mathcal{T}_i$ do not commute with them.

The discrete flow of this set of quantum variables is
\be
 \overline{(q_1,q_2,u,s,Z,a,b)}=(q_1,q_2,u,s,q_2^2Z,ab,b)
\ee
It is easy to check, that this discrete flow preserves the commutation relations \eqref{eq:quantumparameters}.
\begin{conj} \label{th:tauconf}
	Bilinear equations \rf{qPb} for the quantum $\mathcal{T}$-functions are solved in terms of 5d Nekrasov functions
	or $q$-Virasoro conformal blocks:
	\be
	\label{eq:quanumtau}
	\mathcal T_1=a\sum_{m\in\mathbb Z}s^m \mathsf F^{(2)}(uq_2^{4m}|Z),\quad
	\mathcal T_2=ab\sum_{m\in\mathbb Z}s^m \mathsf F^{(2)}(u q_2^{4m}|q_2^2Z),
	\\
	\mathcal T_3=ia\sum_{m\in\frac12+\mathbb Z}s^m \mathsf F^{(2)}(u q_2^{4m}|Z),\quad
	\mathcal T_4=iab\sum_{m\in\mathbb Z+\frac12}s^m \mathsf F^{(2)}(u q_2^{4m}|q_2^2Z).
	\ee
	where the function $\mathsf F$ is defined in the Appendix \ref{App:q-functions}, see \rf{CFdef}.
\end{conj}

This conjecture means that $\mathcal{T}_i$ defined by \eqref{eq:quanumtau} satisfy  the commutation relations \rf{tauLam}, and the bilinear relations \eqref{qPb}. Due to obvious symmetry it is sufficient to check commutations relations between $\mathcal{T}_1, \mathcal{T}_2$; $\mathcal{T}_1, \mathcal{T}_3$; $\mathcal{T}_1, \mathcal{T}_4$ and the first relation in \eqref{qPb}.

By direct calculation, using necessary definitions collected in the Appendix~\ref{App:q-functions}, one can show that these relations turn into bilinear equations in the Conjecture \ref{conj:conf}.
\begin{conj} \label{conj:conf}
	The following bilinear relations hold for 5d Nekrasov functions
	\begin{align}
	\sum_{n \in \mathbb{Z}+\frac14} \left( Z^{2n^2} \mathrm F^{(1)}(uq_1^{4n}|Z)\mathrm F^{(2)}(uq_2^{4n}|Z)\right)= \sum_{n \in \mathbb{Z}-\frac14} \left( Z^{2n^2} \mathrm F^{(1)}(uq_1^{4n}|Z)\mathrm F^{(2)}(uq_2^{4n}|Z)\right), \label{eq:FT1T3}
	\end{align}
	\begin{multline}
	\sum_{n \in \mathbb{Z}\pm \frac14} \left( u^n (q_1q_2)^{2n^2}Z^{2n^2} \mathrm F^{(1)}(uq_1^{4n}|q_1Z)\mathrm F^{(2)}(uq_2^{4n}|q_2Z)\right)=\\ \sum_{n \in \mathbb{Z}\mp \frac14}  \left(u^{-n}(q_1q_2)^{-2n^2}Z^{2n^2}\mathrm F^{(1)}(uq_1^{4n}|q_1^{-1}Z)\mathrm F^{(2)}(uq_2^{4n}|q_2^{-1}Z)\right), \label{eq:FT1T4}
	\end{multline}
	\begin{multline}\label{eq:FT1T2}
	\sum_{2n \in \mathbb{Z}}\left({u^{n}(q_1q_2)^{2n^2}Z^{2n^2}}\mathrm F^{(1)}(uq_1^{4n}|q_1Z)\, \mathrm F^{(2)}(uq_2^{4n}|q_2Z) \right)=\\ \sum_{2n \in \mathbb{Z}}\left(u^{-n}(q_1q_2)^{-2n^2}Z^{2n^2}\mathrm F^{(1)}(uq_1^{4n}|q_1^{-1}Z)\, \mathrm F^{(2)}(uq_2^{4n}|q_2^{-1}Z) \right)
	\end{multline}
	\begin{multline}\label{eq:FT1T1}
	\sum_{2n \in \mathbb{Z}}\left({u^{2n}(q_1q_2)^{4n^2}Z^{2n^2}}\mathrm F^{(1)}(uq_1^{4n}|q_1^2Z)\, \mathrm F^{(2)}(uq_2^{4n}|q_2^{2}Z) \right)=\\= (1-q_1q_2Z)\sum_{2n \in \mathbb{Z}}\left({Z^{2n^2}}\mathrm F^{(1)}(uq_1^{4n}|Z)\, \mathrm F^{(2)}(uq_2^{4n}|Z) \right)
	\end{multline}
\end{conj}
We have checked these relations in first orders in $Z$. The relation \eqref{eq:FT1T1} was proposed in \cite{BS:2016:1}, the conformal $q \rightarrow 1$ limit of the relations \eqref{eq:FT1T3}, \eqref{eq:FT1T2}, \eqref{eq:FT1T1} should follow from the results of \cite{BS:2016:2}. Note that appearance of operator $a$ leads to the fact that bilinear relations
 contain conformal blocks with \emph{different} parameters $q_1,q_2$ (or different central charges).

\begin{Remark}
The formulas \eqref{qPb} give the $\mathcal{T}_i$	in terms of the variables $a,b,u,s,Z,q$. In the formula for $G$ the variables $a,b$ disappear, and $Z,q$ become central. Therefore, one can consider $Z,q$ as numbers and $G=G(Z,q)$ as an operator valued function on them. This operator can act, for example, on the space of functions on $u$, then $s$ becomes the shift operator.

In such approach one can write $\overline{G}=G(qZ,q)$ and consider \eqref{eq:quantumP3G12} as a operator valued difference equation.
\end{Remark}


\section{Discussion}
In this paper we proposed to consider the $q$-Painlev\'e equations as deautonomization of the cluster integrable systems. One of the messages is that quantization of these systems allow to understand better
the intriguing relations between certain Newton polygons, quivers and their mutations, $q$-difference Painlev\'e equations, and 5d supersymmetric gauge theories. Let us finally make few outcoming remarks.
\begin{itemize}
  \item
 For a 5d supersymmetric gauge theory one can assign the SW integrable system, for the gauge group $SU(2)$ always of genus one, see for example \cite[App. A]{Eguchi:2000}. The Newton polygons for these curves, embedded in $\mathbb{C}^\times\times \mathbb{C}^\times$ with natural coordinates, arising from the SW differential, are exactly 4a, 5b, 6b, 7b,  8c from Fig.~\ref{fi:NP} -- for the theories with $N_f=0,1,2,3,4$ correspondingly. On the other hand, we have assigned to these polygons $q$-Painlev\'e equations $A_7^{(1)'}, A_6^{(1)}, A_5^{(1)}, A_4^{(1)}, A_3^{(1)}$, whose tau-functions are expressed in terms of Nekrasov functions for these theories, see Sect.~\ref{ssec:qP:tau}. The polygon $9$  (from Fig \ref{fi:NP}) is the Newton polygon of the 5d $T3$ theory, see \cite[Sec 3.]{Bao:2013}, which corresponds to 5d uplift of the Minahan-Nemeschanski conformal points.

  These 5d gauge theories have extended global symmetry, $E_{r}$ for where $r=N_f+1$ for $5d$ $SU(2)$ theories with $N_f=0,1,2,3,4$ (see \cite{Seib}), and $r=6$ for $T3$ theory. The $q$-Painlev\'e equations we have constructed from corresponding Newton polygons in Sect.~\ref{ss:clqP}, all have the symmetry groups $\tilde{W}(E_{r}^{(1)})$. It looks not like just an accidental coincidence, and it would be interesting to explain appearance of the affinization in this context.

  \item It is already well-known, that quivers arise as effective tool for description of the BPS spectra
  of 4d $\mathcal{N}=2$ supersymmetric gauge theories, see for example \cite{Ceccotti:2011} and references therein. The quivers for $SU(2)$ theories with $N_f$ fundamentals can be found in eq. (2.6) of \cite{Ceccotti:2011}. As was explained above, our quivers $A_7^{(1)'}, A_6^{(1)}, A_5^{(1)}, A_4^{(1)}, A_3^{(1)}$ correspond to the 5d $SU(2)$ gauge theories with $N_f=0,1,2,3,4$. We would like to point out that quivers from \cite{Ceccotti:2011} are obtained from ours by removing of two vertices (for $A_5^{(1)}$ after mutation $\mu_1$, for $A_4^{(1)}$ after mutation $\mu_7$ and for $A_3^{(1)}$ after composition of mutations $\mu_1\circ \mu_2$).

  The quiver corresponding to 5d Minahan-Nemeschanski theory was found in \cite[Sec. 6.3]{Alim:2011}. It is obtained from our quiver $A_2^{(1)}$ after removing one vertex.
We do not know any deep explanation of this fact, but again it does not seem to be just a coincidence.

\item One can ask about generalization of the results of the of the paper to more generic cluster integrable systems (corresponding to polygons different from Fig.~\ref{fi:NP}). It is natural to expect, that deautonomization of these integrable systems leads to the $q$-difference Schlesinger systems.

\item The relation between autonomous limit of Painlev\'e and the SW hamiltonians is known \cite{KMNOY}, \cite{Yamada}, \cite{Mizoguchi}. The fact of appearance of the Newton polygons from Fig.~\ref{fi:NP} in the Painlev\'e theory was also noticed in \cite{Yamada}.

\item In the section \ref{ss:quant} we study quantization of the $q$-Painlev\'e dynamics through their realiziation as compositions of mutations (and permuations). There are different approaches to the quantization, see e.g.  \cite{Hagesawa:2007},\cite{Kuroki},\cite{NY} but in common cases them should be equivalent to our. For example, the formulas for quantum Weyl group actions in  \cite{Hagesawa:2007} are given by conjugation of product of quantum dilogarithms as well as the definition of quantum mutations in \cite{FG:2003}. 

\item The relation between $q$-Painlev\'e and integrable systems was discussed \cite{BGT2}  in the framework of topological string/spectral theory duality. The main proposal is that the tau-function --- analogue of the \eqref{eq:T} --- is equal to the Fredholm determinant $\det(1+\kappa \rho)$. Here analogue means that tau-function in \cite{BGT2} is defined for $|q|=1$ (and not just for $A_7^{(1)}$ surface but for any surface $A_3^{(1)},\ldots,A_8^{(1)}$),  $\kappa$ is an analogue of $u$ in \eqref{eq:T} and in \cite{BGT2} $s=1$ in terms of \eqref{eq:T}. The $\rho$ is inverse of the Hamiltonian of the quantum cluster integrable system, in the simplest example corresponding to \eqref{eq:T} the Hamiltonian of quantum relativistic chain with two particles.

In our terms construction from \cite{BGT2} establishes the relation between quantum autonomous system on one side with the classical non-autonomous system on the other side, corresponding to the same Newton polygon. It would be very interesting to understand this relation better. Another natural question if there is a generalization of this "duality" to our quantum non-autonomus equation.

   \item One more type of Poisson maps between $X$-cluster varieties is given by gluing any two vertices of a quiver (or gluing certain subsets of vertices of two different quivers) and multiplying variables at these vertices, with adding corresponding edges taking into account their orientations or directions of the arrows~--- e.g. two opposite arrows should be annihilated. It is easy to see, for example, that all quivers from Fig.~\ref{fi:PQ} can be obtained consequently by gluing a pair of vertices according to the Fig.~\ref{fi:qPeq} (more precisely for $A_2^{(1)}$ quiver one gets quiver obtained from $A_3^{(1)}$ quiver by one mutation).

  \item The remaining two $q$-Painlev\'e equations with the symmetry types $A_0^{(1)}$ and  $A_1^{(1)}$ can be also obtained from mutations of the $X$-cluster varieties. Corresponding quivers are presented in Fig.~\ref{fi:PQA01}.

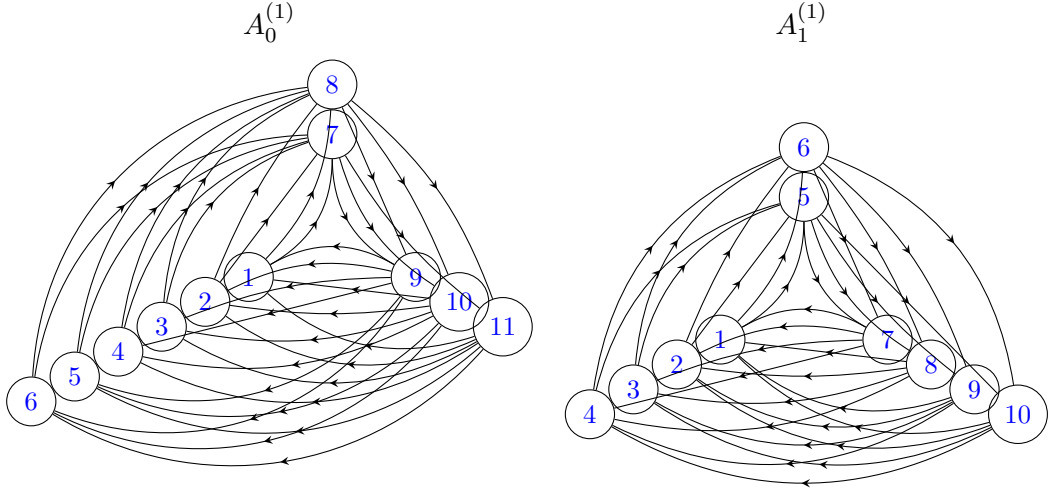
\begin{figure}[h]
\begin{center}
	\begin{tabular}{c c}
  	$A_0^{(1)}$ & $A_1^{(1)}$  \vspace{0.2cm}
  	\\
   	\begin{tikzpicture}[scale=2.2, font = \small]
		 \node[shape=circle,draw=black] (A) at (-0.5,0) {\color{blue}1};
		 \node[shape=circle,draw=black] (A1) at (-0.76,-0.15) {\color{blue}2};
		 \node[shape=circle,draw=black] (A2) at (-1.02,-0.30) {\color{blue}3};
		 \node[shape=circle,draw=black] (A3) at (-1.28,-0.45) {\color{blue}4};
 		 \node[shape=circle,draw=black] (A4) at (-1.54,-0.60) {\color{blue}5};
		 \node[shape=circle,draw=black] (A5) at (-1.80,-0.75) {\color{blue}6};
		 \node[shape=circle,draw=black] (B) at (0.5,0) {\color{blue}9};
		 \node[shape=circle,draw=black] (B1) at (0.76,-0.15) {\color{blue}10};
		 \node[shape=circle,draw=black] (B2) at (1.02,-0.30) {\color{blue}11};	
		 \node[shape=circle,draw=black]  (C) at (0,0.86) {\color{blue}7};
		 \node[shape=circle,draw=black] (C1) at (0,1.16) {\color{blue}8};
		 \path
		 \foreach \from/\to in {C1/B,B1/A,A1/C,B/A3}
		 {
		 	(\from) edge[mid arrow] (\to)					
		 }
		 \foreach \from/\to in {B2/A,B2/A1,B2/A2,B2/A3,A2/C,A2/C1,A3/C1,A3/C}
		 {
		 	(\from) edge[mid arrow,bend left=30] (\to)					
		 }
		 \foreach \from/\to in {C/B,B/A,A/C}
		 {
		 	(\from) edge[mid arrow,bend right=30] (\to)					
		 }
		 \foreach \from/\to in {C/B1,B/A1,A/C1}
		 {
		 	(\from) edge[mid arrow,bend right=20] (\to)					
		 }
		 \foreach \from/\to in {C/B2,B/A2}
		 {
		 	(\from) edge[mid arrow,bend right=10] (\to)					
		 }
		 \foreach \from/\to in {B1/A3}
		 {
		 	(\from) edge[mid arrow,bend left=20] (\to)					
		 }
		 \foreach \from/\to in {C1/B2,B1/A2}
		 {
		 	(\from) edge[mid arrow,bend left=15] (\to)					
		 }
		 \foreach \from/\to in {C1/B1,B1/A1,A1/C1}
		 {
		 	(\from) edge[mid arrow,bend left=8] (\to)					
		 }
		\foreach \from/\to in
		{A4/C1,A4/C,B/A4,B1/A4,B2/A4}
		{
		(\from) edge[mid arrow,bend left=35] (\to)					
		}	
		\foreach \from/\to in
		{A5/C1,A5/C,B/A5,B1/A5,B2/A5}
		{
			(\from) edge[mid arrow,bend left=40] (\to)					
		}	
		;
	\end{tikzpicture} 	
  	&
  	\begin{tikzpicture}[scale=2.2, font = \small]
  	\node[shape=circle,draw=black] (A) at (-0.5,0) {\color{blue}1};
  	\node[shape=circle,draw=black] (A1) at (-0.76,-0.15) {\color{blue}2};
  	\node[shape=circle,draw=black] (A2) at (-1.02,-0.30) {\color{blue}3};
	\node[shape=circle,draw=black] (A3) at (-1.28,-0.45) {\color{blue}4};
	\node[shape=circle,draw=black] (B) at (0.5,0) {\color{blue}7};
  	\node[shape=circle,draw=black] (B1) at (0.76,-0.15) {\color{blue}8};
  	\node[shape=circle,draw=black] (B2) at (1.02,-0.30) {\color{blue}9};	
   	\node[shape=circle,draw=black] (B3) at (1.28,-0.45) {\color{blue}10};	
   	\node[shape=circle,draw=black]  (C) at (0,0.86) {\color{blue}5};
  	\node[shape=circle,draw=black] (C1) at (0,1.16) {\color{blue}6};
  	\path
  	\foreach \from/\to in {C1/B,B1/A,A1/C,C/B3,B/A3}
  	{
  		(\from) edge[mid arrow] (\to)					
  	}
  	\foreach \from/\to in {B2/A,B2/A1,B2/A2,B2/A3,A2/C,A2/C1,B3/A3,A3/C1,C1/B3,A3/C,B3/A2,B3/A1,B3/A}
  	{
  		(\from) edge[mid arrow,bend left=30] (\to)					
  	}
  	\foreach \from/\to in {C/B,B/A,A/C}
  	{
  		(\from) edge[mid arrow,bend right=30] (\to)					
  	}
  	\foreach \from/\to in {C/B1,B/A1,A/C1}
  	{
  		(\from) edge[mid arrow,bend right=20] (\to)					
  	}
  	\foreach \from/\to in {C/B2,B/A2}
  	{
  		(\from) edge[mid arrow,bend right=10] (\to)					
  	}
	\foreach \from/\to in {B1/A3}
	{
		(\from) edge[mid arrow,bend left=20] (\to)					
	}
  	\foreach \from/\to in {C1/B2,B1/A2}
  	{
  		(\from) edge[mid arrow,bend left=15] (\to)					
  	}
  	\foreach \from/\to in {C1/B1,B1/A1,A1/C1}
  	{
  		(\from) edge[mid arrow,bend left=8] (\to)					
  	}
  	;
  	\end{tikzpicture}
  \end{tabular}
  \caption{Quivers of $A_0^{(0)}$ and $A_1^{(1)}$ $q$-Painlev'e equations}
  \label{fi:PQA01}
\end{center}
\end{figure}

  The generators of the group $\tilde{\mathcal{G}}_\mathcal{Q}$ are constructed here similarly to the case $A_2^{(1)}$, the resulting formulas can be identified with those of \cite{Tsuda:2006:2}. We do not know yet how to obtain these quivers from (a version of?) approach of \cite{GK:2011,FM:2014}.

  \item In this paper we restrict ourselves to the $q$-Painlev\'e case, but there should exist an analogue of the story in the case of quantum \emph{differential} Painlev\'e. It seems that these formulas for quantum tau-functions could be obtained as a limit $q\rightarrow 1$ of  \eqref{eq:quanumtau},
since the corresponding bilinear relations on conformal blocks were obtained in \cite{BS:2014}.

%

	\item Quivers $A_7^{(1)}$ and $A_6^{(1)}$ are called 4-Somos and 5-Somos quivers correspondingly due to their relation to the Somos sequences \cite{Okubo:2017}.
\end{itemize}


\section*{Acknowledgements}
We would like to thank G.~Bonelli, M.~Caorsi, A.~Dzhamay, B.~Feigin, V.~Fock, A.~Grassi, A.~Hanany, O.~Lisovyy, A.~Pogrebkov, A.~Shchechkin, R.-K. Seong,  M.~Semenyakin, N.~Sopenko, A.~Tanzini, T.~Tsuda, Y.~Yamada  for stimulating discussions. We are especially grateful to A.~Shchechkin for careful reading of preliminary version of the paper and many critical remarks.

We would like to thank the KdV Institute of
the University of Amsterdam, where this work has  been started within the framework of Student workshop ``Representation Theory and Integrable Systems'' in May 2017, and especially E.~Opdam and G.~Helminck, who organized this meeting. MB and PG are also indebted to the University of Tours, and especially 
to O.~Lisovyy, for the hospitality during their visit, where some preliminary ideas of this work have been first discussed. MB is grateful to SISSA, and especially 
to A.~Tanzini and G.~Bonelli for the hospitality during his visit to Trieste, where a part of this work has been done. We are also grateful to the organizers of the workshop \textit{Topological Field Theories, String theory and Matrix Models} in August 2017 for the opportunity to present and discuss our preliminary results.

This work has been funded by the Russian Academic Excellence Project `5-100'. The main results of Sect.~\ref{ssec:clqPX} and Sect.~\ref{ssec:quantF} have been obtained using support of Russian Science Foundation under the grant No. 16-11-10160. The work of AM has been also partially supported by RFBR grant 17-01-00585. MB and PG are the \emph{Young Russian Mathematics} award winners and would like to thank its sponsors and jury.


\appendix
\section{Newton polygons, bipartite graphs and quivers} \label{ap:catalog}

\paragraph{Polygon 3}
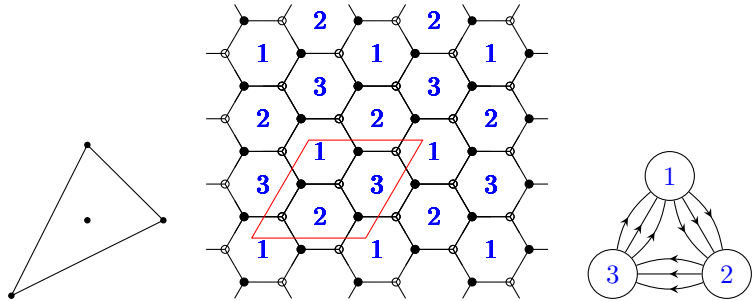
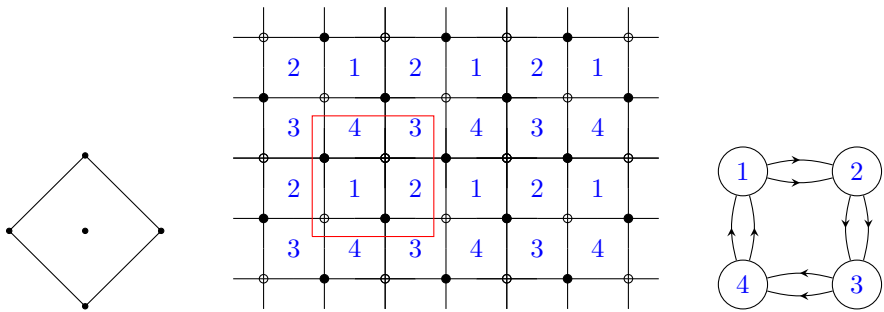
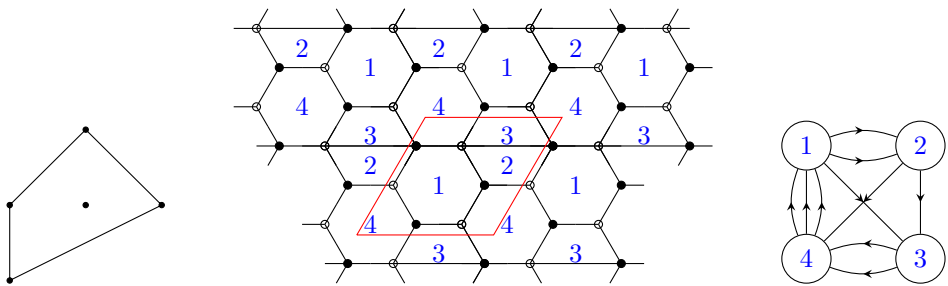
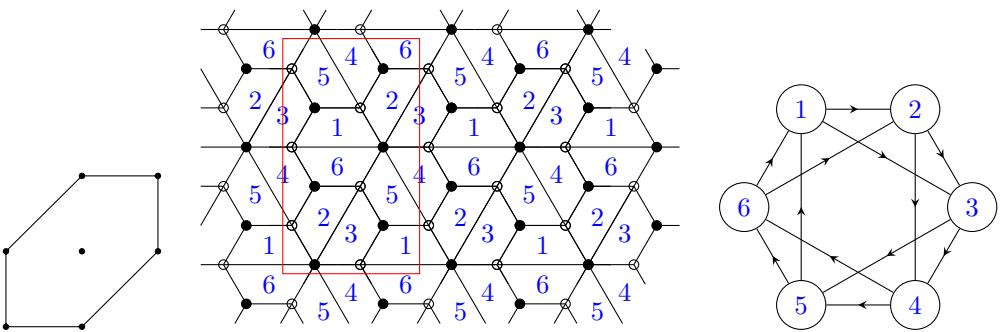
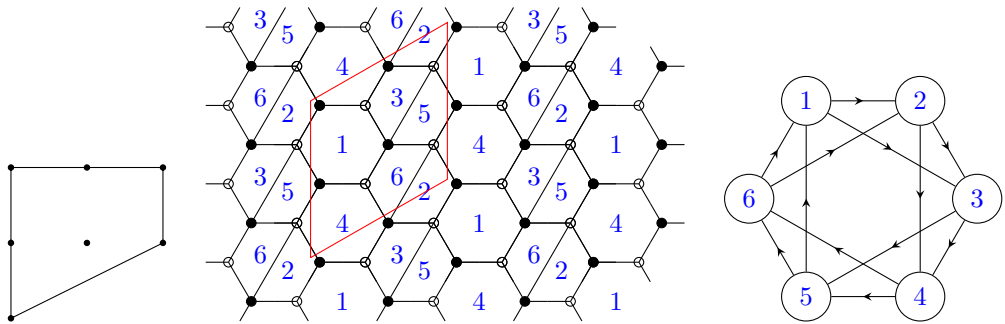
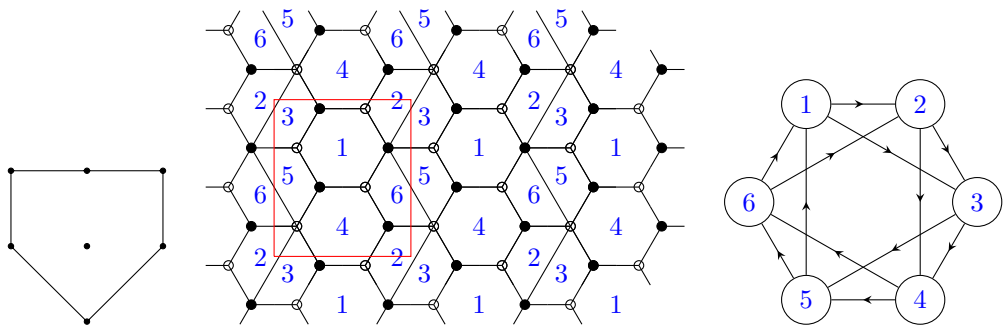
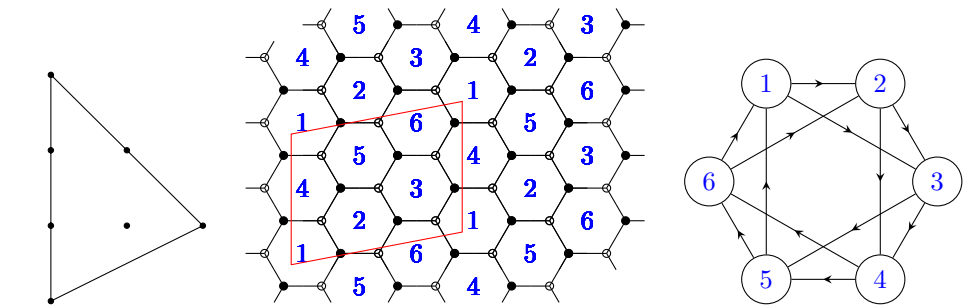
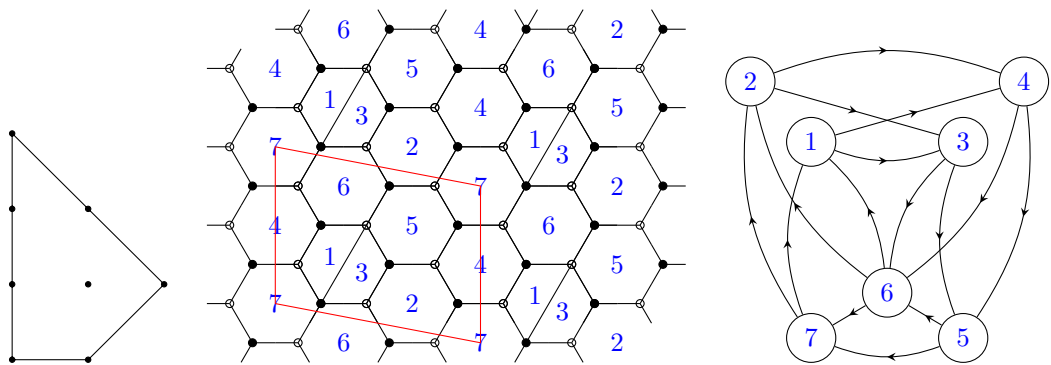
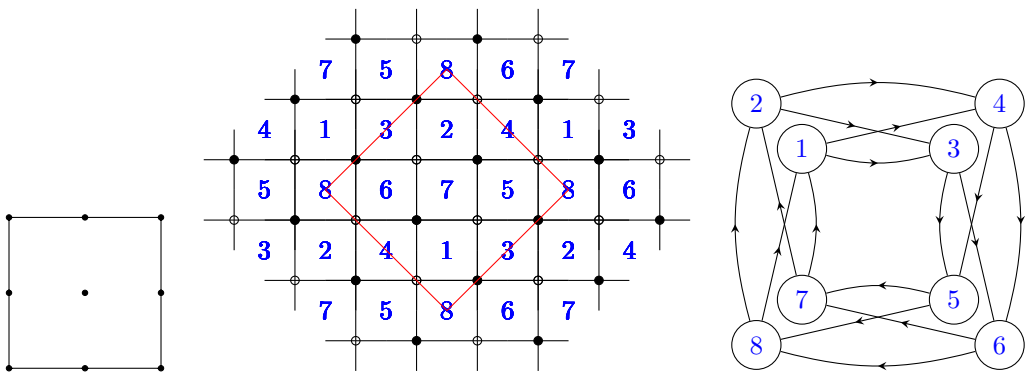
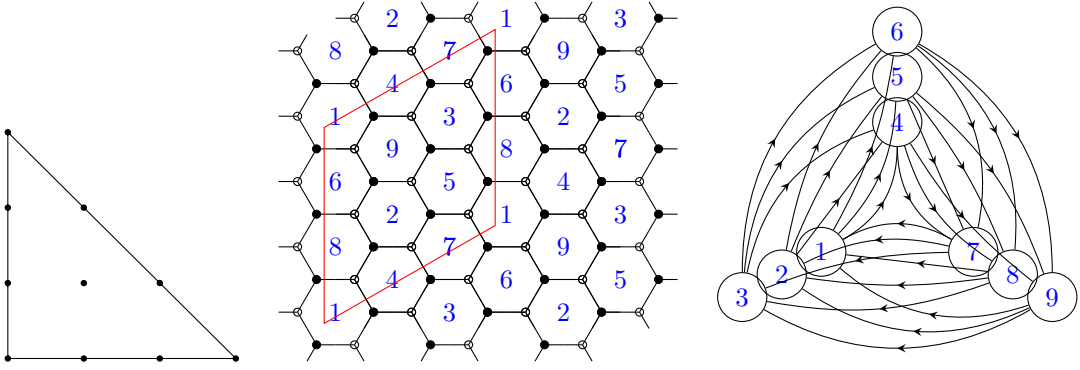
\begin{center}
	\begin{tikzpicture}[scale=1, font = \small]
\draw[fill] (0,1) circle (1pt) -- (1,0) circle (1pt) -- (-1,-1) circle (1pt) -- (0,1);
\draw[fill] (0,0) circle (1pt);
\end{tikzpicture}
\quad
\begin{tikzpicture}[scale=0.5, font = \small]
\foreach \i/\j in {-3/6,-2/4,-2/5,-2/6,-1/2,-1/3,-1/4,-1/5,-1/6,0/0,0/1,0/2,0/3,0/4,1/0,1/1,1/2,2/0}
{
	\coordinate (period) at (3*\i+1.5*\j,sin{60}*\j);	
	\foreach \pointb in {(1,0), (-0.5,-sin{60}), (-0.5,sin{60})}
	{
		\draw[fill] \pointb ++(period) circle (3pt);
		\foreach \a in {0,120,-120} \draw[thin] \pointb ++(period) -- +(\a:0.5);
	}
	\foreach \pointw in {(0.5,sin{60}), (0.5,-sin{60}),  (-1,0)}
	{
		\draw \pointw ++(period) circle (3pt);
		\foreach \a in {60,180,-60} \draw[thin] \pointw ++(period) -- +(\a:0.5);
	}
	\foreach \point in {(0,0),(0,6*sin{60}),(4.5,3*sin{60}),(1.5,3*sin{60}),(3,6*sin{60}),(3,0),(6,0),(6,6*sin{60})} {\node at \point {\color{blue}1};}
	\foreach \point in {(1.5,sin{60}),(1.5,7*sin{60}),(3,4*sin{60}),(6,4*sin{60}),(0,4*sin{60}),(4.5,7*sin{60}),(4.5,1*sin{60})} {\node at \point {\color{blue}2};}
	\foreach \point in {(0,2*sin{60}),(4.5,5*sin{60}),(3,2*sin{60}),(3,2*sin{60}),(1.5,5*sin{60}),(6,2*sin{60})} {\node at \point {\color{blue}3};}
}
\draw[color=red] (-0.3,0.3) -- +(3,0) -- +(4.5,3*sin{60}) -- +(1.5,3*sin{60}) -- +(0,0);
\end{tikzpicture}
\quad 	
	\begin{tikzpicture}[scale=1.5, font = \small]
	\node[shape=circle,draw=black] (C) at (-0.5,0) {\color{blue}3};
	\node[shape=circle,draw=black] (B) at (0.5,0) {\color{blue}2};
	\node[shape=circle,draw=black] (A) at (0,cos{30}) {\color{blue}1};
	\path
	\foreach \from/\to in {A/B,B/C,C/A}
	{
		(\from) edge[mid arrow,bend right=20] (\to)
		(\from) edge[mid arrow,bend left=20] (\to)
		(\from) edge[mid arrow] (\to)
	};			
	\end{tikzpicture} 		
\end{center}

\paragraph{Polygon 4a}
\begin{center}
 	\begin{tikzpicture}[scale=1, font = \small]
 \draw[fill] (0,1) circle (1pt) -- (1,0) circle (1pt) -- (0,-1) circle (1pt) -- (-1,0) circle (1pt) -- (0,1);
 \draw[fill] (0,0) circle (1pt);
 \end{tikzpicture}
 		\qquad
\begin{tikzpicture}[scale=0.8, font = \small]
	\foreach \i in {0,1,2}
	\foreach \j in {0,1}
	{
	\coordinate (period) at (3*\i+1.5*\j,3*sin{60}*\j);	
	\foreach \pointb in {(0.5,-0.5), (-0.5,0.5),  (-0.5,-1.5), (-1.5,-0.5)}
	{
		\draw[fill] \pointb ++(2*\i,2*\j) circle (2pt);
		\foreach \a in {0,90,180,270} \draw[thin] \pointb ++(2*\i,2*\j) -- +(\a:0.5);
	}
	\foreach \pointw in {(0.5,0.5), (0.5,-1.5),  (-0.5,-0.5), (-1.5,0.5), (-1.5,-1.5)}
	{
		\draw \pointw ++(2*\i,2*\j) circle (2pt);
		\foreach \a in {0,90,180,270} \draw[thin] \pointw ++(2*\i,2*\j) -- +(\a:0.5);
	}
	\node at (2*\i,2*\j) 	{\color{blue}1};
	\draw (2*\i,2*\j) ++(0,-1) node {\color{blue}4};
	\draw (2*\i,2*\j) ++(-1, 0) node {\color{blue}2};
	\draw (2*\i,2*\j) ++(-1, -1) node {\color{blue}3};
	}
	\draw[color=red] (-0.7,-0.8) -- +(2,0) -- +(2,2) -- +(0,2) -- +(0,0);
\end{tikzpicture}\qquad
\begin{tikzpicture}[scale=1.5, font = \small]
	\node[shape=circle,draw=black] (A) at (0,1) {\color{blue}1};
	\node[shape=circle,draw=black] (B) at (1,1) {\color{blue}2};
	\node[shape=circle,draw=black] (C) at (1,0) {\color{blue}3};
	\node[shape=circle,draw=black] (D) at (0,0) {\color{blue}4};
	\path \foreach \from/\to in {A/B,B/C,C/D,D/A}
	{
		(\from) edge[mid arrow,bend right=15] (\to)
		(\from) edge[mid arrow,bend left=15] (\to)
	};			
\end{tikzpicture}
\end{center}

\paragraph{Polygon 4b}
\begin{center}
	\begin{tikzpicture}[scale=1, font = \small]
		\draw[fill] (0,1) circle (1pt) -- (1,0) circle (1pt) -- (-1,-1) circle (1pt) -- (-1,0) circle (1pt) -- (0,1);
		\draw[fill] (0,0) circle (1pt);
	\end{tikzpicture}
	\qquad
	\begin{tikzpicture}[scale=0.6, font = \small]
	\foreach \i/\j in {0/0, 1/0,  0/1,-1/1,1/1}
	{
		\coordinate (period) at (3*\i+1.5*\j,3*sin{60}*\j);
		\foreach \pointb in {(1,0), (1,-2*sin{60}),  (-0.5,sin{60}), (-0.5,-sin{60}), (-2,0), (-2,-2*sin{60})}
		{
			\draw[fill] \pointb ++(period) circle (2.5pt);
			\foreach \a in {0,120,-120} \draw[thin] \pointb ++(period) -- +(\a:0.5);
		}
		\foreach \pointw in {(-1,0), (-1,-2*sin{60}), (0.5,sin{60}), (0.5,-sin{60}), (-2.5,sin{60}), (-2.5,-sin{60})}
		{
			\draw \pointw ++(period) circle (2.5pt);
			\foreach \a in {60,180,-60} \draw[thin] \pointw ++(period) -- +(\a:0.5);
		}
		\draw[thin] (-2.5,sin{60}) ++(period) -- +(0:2);		
		\draw[thin] (-1,-2*sin{60}) ++(period) -- +(0:2);
		\draw[thin] (-2,-2*sin{60}) ++(period) -- +(180:0.5);
		\draw[thin] (0.5,sin{60}) ++(period) -- +(0:0.5);							
		\node at (period) 	{\color{blue}1};
		\draw (period) ++(-1.5,-sin{60}) node {\color{blue}4};
		\draw (period) ++(-1.5,0.5*sin{60}) node {\color{blue}2};
		\draw (period) ++(0,-1.5) node {\color{blue}3};		
	}	
	\draw[color=red] (-1.8,-1.1) -- +(3,0) -- +(4.5,3*sin{60}) -- +(1.5,3*sin{60}) -- +(0,0);
\end{tikzpicture}
\qquad
\begin{tikzpicture}[scale=1.5, font = \small]
	\node[shape=circle,draw=black] (A) at (0,1) {\color{blue}1};
	\node[shape=circle,draw=black] (B) at (1,1) {\color{blue}2};
	\node[shape=circle,draw=black] (C) at (1,0) {\color{blue}3};
	\node[shape=circle,draw=black] (D) at (0,0) {\color{blue}4};
	\path
	\foreach \from/\to in {A/B,C/D,D/A}
	{
		(\from) edge[mid arrow,bend right=20] (\to)
		(\from) edge[mid arrow,bend left=20] (\to)
	};			
	\path
	\foreach \from/\to in {B/C,B/D,D/A,A/C}
	{
		(\from) edge[mid arrow] (\to)
	};			
\end{tikzpicture}
\end{center}

\paragraph{Polygon 4c}
\begin{center}
	\begin{tikzpicture}[scale=1, font = \small]
		\draw[fill] (0,1) circle (1pt) -- (1,0) circle (1pt) -- (-2,-1) circle (1pt) -- (-1,0) circle (1pt) -- (0,1);
		\draw[fill] (0,0) circle (1pt);
	\end{tikzpicture}
	\qquad
	\begin{tikzpicture}[scale=0.5, font = \small]
	\foreach \i in {0,1}
	\foreach \j in {0,1}
	{
		\coordinate (period) at (3*\i,4*sin{60}*\j);	
		\foreach \pointb in {(1,0), (1,-2*sin{60}), (-0.5,sin{60}), (-0.5,-sin{60}), (-2,0), (-2,-2*sin{60})}
		{
			\draw[fill] \pointb ++(period) circle (3pt);
			\foreach \a in {0,120,-120} \draw[thin] \pointb ++(period) -- +(\a:0.5);
		}
		\foreach \pointw in {(-1,0), (-1,-2*sin{60}),  (0.5,sin{60}), (0.5,-sin{60}), (-2.5,sin{60}), (-2.5,-sin{60})}
		{
			\draw \pointw ++(period) circle (3pt);
			\foreach \a in {60,180,-60} \draw[thin] \pointw ++(period) -- +(\a:0.5);
		}		
		\node at (period) 	{\color{blue}3};
		\draw (period) ++(0,-2*sin{60}) node {\color{blue}1};
		\draw (period) ++(-1.5, -sin{60}) node {\color{blue}2};
		\draw (period) ++(-1.5, +sin{60}) node {\color{blue}4};		
	}	
	\draw[color=red] (-1.6,0.3) -- +(3,0) -- +(3,4*sin{60}) -- +(0,4*sin{60}) -- +(0,0);
	\end{tikzpicture}
	\qquad
\begin{tikzpicture}[scale=1.5, font = \small]
	\node[shape=circle,draw=black] (A) at (0,1) {\color{blue}1};
	\node[shape=circle,draw=black] (B) at (1,1) {\color{blue}2};
	\node[shape=circle,draw=black] (C) at (1,0) {\color{blue}3};
	\node[shape=circle,draw=black] (D) at (0,0) {\color{blue}4};
	\path \foreach \from/\to in {A/B,B/C,C/D,D/A}
	{
		(\from) edge[mid arrow,bend right=15] (\to)
		(\from) edge[mid arrow,bend left=15] (\to)
	};			
\end{tikzpicture}
\end{center}

\paragraph{Polygon 5a}
\begin{center}
\begin{tikzpicture}[scale=1, font = \small]
	\draw[fill] (0,1) circle (1pt) -- (1,1) circle (1pt) -- (1,0) circle (1pt) -- (0,-1) circle (1pt) -- (-1,0) circle (1pt) -- (0,1);
	\draw[fill] (0,0) circle (1pt);
\end{tikzpicture}
\qquad
\begin{tikzpicture}[scale=0.6, font = \small]
	\foreach \i/\j in {0/0,0/1,1/0,1/1}
	{
	\coordinate (period) at (3*\i+1.5*\j,3*sin{60}*\j);	
	\foreach \pointb in {(1,0), (2.5,sin{60}), (-0.5,sin{60}), (-0.5,-sin{60}), (1,-2*sin{60}),  (-2,-2*sin{60})}
	{
		\draw[fill] \pointb ++(period) circle (2.5pt);
		\foreach \a in {0,120,-120} \draw[thin] \pointb ++(period) -- +(\a:0.5);
	}
	\foreach \pointw in {(2,0), (0.5,sin{60}), (-1,-2*sin{60}), (0.5,-sin{60}), (-1,0)}
	{
		\draw \pointw ++(period) circle (2.5pt);
		\foreach \a in {60,180,-60} \draw[thin] \pointw ++(period) -- +(\a:0.5);
	}
	\draw[thin] (2.5,sin{60}) ++(period) -- +(60:0.5);
	\draw[thin] (-0.5,sin{60}) ++(period) -- +(60:0.5);
	\draw[thin] (-0.5,sin{60}) ++(period) -- +(180:0.5);
	\draw[thin] (-2,-2*sin{60}) ++(period) -- +(180:0.5);
	\draw[thin] (0.5,sin{60}) ++(period) -- +(0:2);			
	\draw[thin] (-1,-2*sin{60}) ++(period) -- +(0:2);			
	\draw[thin] (1,-2*sin{60}) ++(period) -- +(60:2);			
	\draw[thin] (-2,-2*sin{60}) ++(period) -- +(60:2);		
	\node at (period) {\color{blue}2};
	\draw (period) ++(1.5,0.5*sin{60}) node {\color{blue}3};
	\draw (period) ++(0,-1.5*sin{60}) node {\color{blue}4};
	\draw (period) ++(1.2,-0.9*sin{60}) node {\color{blue}5};
	\draw (period) ++(-1.2,-1.2*sin{60}) node {\color{blue}1};
	}
	\draw[color=red] (0.8,-1.1) -- +(3,0) -- +(4.5,3*sin{60}) -- +(1.5,3*sin{60}) -- +(0,0);
\end{tikzpicture}
\qquad
\begin{tikzpicture}[scale=1.5, font = \small]
	\node[shape=circle,draw=black] (A) at (0,1.62) {\color{blue}2};
	\node[shape=circle,draw=black] (B) at (0.8,0.95) {\color{blue}3};
	\node[shape=circle,draw=black] (C) at (0.5,0) {\color{blue}4};
	\node[shape=circle,draw=black] (D) at (-0.5,0) {\color{blue}5};
	\node[shape=circle,draw=black] (E) at (-0.8,0.95) {\color{blue}1};
	\path
	\foreach \from/\to in {A/B,E/A}
	{
		(\from) edge[mid arrow,bend right=15] (\to)
		(\from) edge[mid arrow,bend left=15] (\to)
	}
	\foreach \from/\to in {B/C,B/D,C/E,D/E,C/D,A/C,D/A}
	{
		(\from) edge[mid arrow] (\to)					
	}
	;				
\end{tikzpicture} 						
\end{center}

\paragraph{Polygon 5b}
\begin{center}
\begin{tikzpicture}[scale=1, font = \small]
	\draw[fill] (0,1) circle (1pt) -- (1,0) circle (1pt) -- (-1,-1) circle (1pt) -- (-1,0) circle (1pt) -- (-1,1) circle (1pt) -- (0,1);
	\draw[fill] (0,0) circle (1pt);	
\end{tikzpicture}
\quad
\begin{tikzpicture}[scale=0.6, font = \small]
\foreach \i/\j in {0/0,1/0,0/1,1/1}
{
	\coordinate (period) at (3*\i,4*sin{60}*\j);	
	\foreach \pointb in {(1,0), (1,-2*sin{60}), (-0.5,sin{60}), (-0.5,-sin{60}), (-0.5,-3*sin{60}), (-2,0),  (-2,-2*sin{60})}
	{
		\draw[fill] \pointb ++(period) circle (2.5pt);
		\foreach \a in {0,120,-120} \draw[thin] \pointb ++(period) -- +(\a:0.5);
	}
	\foreach \pointw in {(0.5,sin{60}), (0.5,-sin{60}), (0.5,-3*sin{60}), (-1,0), (-1,-2*sin{60}), (-2.5,sin{60}), (-2.5,-sin{60})}
	{
		\draw \pointw ++(period) circle (2.5pt);
		\foreach \a in {60,180,-60} \draw[thin] \pointw ++(period) -- +(\a:0.5);
	}
	\draw[thin] (-0.5,-3*sin{60}) ++(period) -- +(60:2);		
	\node at (period) {\color{blue}2};
	\draw (-1.5,-sin{60}) ++(period) node {\color{blue}1};
	\draw (-1.5,sin{60}) ++(period) node {\color{blue}3};
	\draw (-0.4,-1.8*sin{60}) ++(period) node {\color{blue}4};
	\draw (0.5,-2.1*sin{60}) ++(period) node {\color{blue}5};
}
\draw[color=red] (-1.8,-1.1) -- +(3,0) -- +(3,4*sin{60}) -- +(0,4*sin{60}) -- +(0,0);
\end{tikzpicture}
\quad
\begin{tikzpicture}[scale=1.5, font = \small]
	\node[shape=circle,draw=black] (A) at (0,1.62) {\color{blue}2};
	\node[shape=circle,draw=black] (B) at (0.8,0.95) {\color{blue}3};
	\node[shape=circle,draw=black] (C) at (0.5,0) {\color{blue}4};
	\node[shape=circle,draw=black] (D) at (-0.5,0) {\color{blue}5};
	\node[shape=circle,draw=black] (E) at (-0.8,0.95) {\color{blue}1};
	\path
	\foreach \from/\to in {A/B,E/A}
	{
		(\from) edge[mid arrow,bend right=15] (\to)
		(\from) edge[mid arrow,bend left=15] (\to)
	}
	\foreach \from/\to in {B/C,B/D,C/E,D/E,C/D,A/C,D/A}
	{
		(\from) edge[mid arrow] (\to)					
	}
	;				
\end{tikzpicture} 						
\end{center}

\paragraph{Polygon 6a}
\begin{center}
\begin{tikzpicture}[scale=1, font = \small]
	\draw[fill] (0,1) circle (1pt) -- (1,1) circle (1pt) -- (1,0) circle (1pt) -- (0,-1) circle (1pt) -- (-1,-1) circle (1pt) -- (-1,0) circle (1pt) -- (0,1);
	\draw[fill] (0,0) circle (1pt);
\end{tikzpicture}
\quad
\begin{tikzpicture}[scale=0.6, font = \small]
\foreach \i/\j in {-3/6,-2/4,-2/5,-2/6,-1/2,-1/3,-1/4,-1/5,-1/6,0/0,0/1,0/2,0/3,0/4,0/5,1/0,1/1,1/2,1/3,2/0,2/1}
{
	\coordinate (period) at (3*\i+1.5*\j,sin{60}*\j);	
	\foreach \pointb in {(1,0), (-0.5,-sin{60})}
	{
		\draw[fill] \pointb ++(period) circle (3pt);
		\foreach \a in {0,120,-120} \draw[thin] \pointb ++(period) -- +(\a:0.5);
	}
	\foreach \pointw in {(0.5,-sin{60}),  (-1,0)}
	{
		\draw \pointw ++(period) circle (3pt);
		\foreach \a in {60,180,-60} \draw[thin] \pointw ++(period) -- +(\a:0.5);
	}
}
\foreach \i/\j in {-0.5/3*sin{60},2.5/3*sin{60},5.5/3*sin{60},1/0*sin{60},4/0*sin{60},7/0*sin{60}} \draw[thin] (\i,\j) -- +(60:2);
\foreach \i/\j in {-0.5/3*sin{60},2.5/3*sin{60},5.5/3*sin{60},1/0*sin{60},4/0*sin{60},7/0*sin{60}} \draw[thin] (\i,\j) -- +(60:2);
\foreach \i/\j in {2.5/3*sin{60},5.5/3*sin{60},8.5/3*sin{60},1/0*sin{60},4/0*sin{60},7/0*sin{60},1/6*sin{60},4/6*sin{60},7/6*sin{60}} \draw[thin] (\i,\j) -- +(180:2);
\draw[thin] (-0.5,3*sin{60}) -- +(180:1);
\foreach \i/\j in {-0.5/3*sin{60},2.5/3*sin{60},5.5/3*sin{60},1/6*sin{60},4/6*sin{60},7/6*sin{60}} \draw[thin] (\i,\j) -- +(300:2);
\foreach \i/\j in {1/0*sin{60},4/0*sin{60},7/0*sin{60}} \draw[thin] (\i,\j) -- +(300:1.5);
\foreach \i/\j in {1/6*sin{60},4/6*sin{60},7/6*sin{60}} \draw[thin] (\i,\j) -- +(60:0.5);
\foreach \i/\j in {0.5/-sin{60},3.5/-sin{60},6.5/-sin{60}} \draw (\i,\j) -- +(240:0.5);
\foreach \i/\j in {-1/2*sin{60}} \draw (\i,\j) -- +(240:1);
\foreach \i/\j in {-1/4*sin{60}} \draw (\i,\j) -- +(120:1);
\foreach \i/\j in {2.5/sin{60},5.5/sin{60},8.5/sin{60},1/4*sin{60},4/4*sin{60},7/4*sin{60}} 	
{
	\draw (\i,\j) ++(-0.7,-0.2*sin{60}) node {\color{blue}3};
	\draw (\i,\j) ++(-1.3,0.2*sin{60}) node {\color{blue}2};
}
\foreach \i/\j in {1/2*sin{60},4/2*sin{60},7/2*sin{60},2.5/5*sin{60},5.5/5*sin{60},8.5/5*sin{60},2.5/-1*sin{60},5.5/-1*sin{60},8.5/-1*sin{60}} 	 
{
	\draw (\i,\j) ++(-1.3,-0.2*sin{60}) node {\color{blue}5};
	\draw (\i,\j) ++(-0.7,0.3*sin{60}) node {\color{blue}4};	
}
\foreach \i/\j in {0/0,3/0,6/0,1.5/3*sin{60},4.5/3*sin{60},7.5/3*sin{60}} 	
{
	\draw (\i,\j) ++(0,-0.5*sin{60}) node {\color{blue}6};
	\draw (\i,\j) ++(0,0.5*sin{60}) node {\color{blue}1};
}
\foreach \i/\j in {0/6*sin{60},3/6*sin{60},6/6*sin{60}} 	
{
	\draw (\i,\j) ++(0,-0.5*sin{60}) node {\color{blue}6};
}
\draw[color=red] (0.3,-0.2) -- +(3,0) -- +(3,6*sin{60}) -- +(0,6*sin{60}) -- +(0,0);
\end{tikzpicture}
\quad	
\begin{tikzpicture}[scale=1.5, font = \small]
	\node[shape=circle,draw=black] (A) at (0,1.73) {\color{blue}1};
	\node[shape=circle,draw=black] (B) at (1,1.73) {\color{blue}2};
	\node[shape=circle,draw=black] (C) at (1.5,0.865) {\color{blue}3};
	\node[shape=circle,draw=black] (D) at (1,0) {\color{blue}4};
	\node[shape=circle,draw=black] (E) at (0,0) {\color{blue}5};
	\node[shape=circle,draw=black] (F) at (-0.5,0.865) {\color{blue}6};
	\path
	\foreach \from/\to in {A/B,B/C,C/D,D/E,E/F,F/A,A/C,C/E,E/A,B/D,D/F,F/B}
	{
		(\from) edge[mid arrow] (\to)					
	}
	;				
\end{tikzpicture} 		
\end{center}

\newpage
\paragraph{Polygon 6b}
\begin{center}
\begin{tikzpicture}[scale=1, font = \small]
	\draw[fill] (0,1) circle (1pt) -- (1,1) circle (1pt) -- (1,0) circle (1pt) -- (-1,-1) circle (1pt) -- (-1,0) circle (1pt) -- (-1,1) circle (1pt) -- (0,1);
	\draw[fill] (0,0) circle (1pt);
\end{tikzpicture}
\quad
\begin{tikzpicture}[scale=0.6, font = \small]
\foreach \i/\j in {-3/6,-2/4,-2/5,-2/6,-1/2,-1/3,-1/4,-1/5,-1/6,0/0,0/1,0/2,0/3,0/4,0/5,1/0,1/1,1/2,1/3,2/0,2/1}
{
	\coordinate (period) at (3*\i+1.5*\j,sin{60}*\j);	
	\foreach \pointb in {(1,0), (-0.5,-sin{60})}
	{
		\draw[fill] \pointb ++(period) circle (3pt);
		\foreach \a in {0,120,-120} \draw[thin] \pointb ++(period) -- +(\a:0.5);
	}
	\foreach \pointw in {(0.5,-sin{60}),  (-1,0)}
	{
		\draw \pointw ++(period) circle (3pt);
		\foreach \a in {60,180,-60} \draw[thin] \pointw ++(period) -- +(\a:0.5);
	}
}
\foreach \i/\j in {-0.5/3*sin{60},-0.5/-sin{60},2.5/3*sin{60},2.5/-sin{60},5.5/3*sin{60},5.5/-sin{60},-0.5/1*sin{60},2.5/1*sin{60},5.5/1*sin{60}} \draw[thin] (\i,\j) -- +(60:2);
\foreach \i/\j in {-0.5/5*sin{60},2.5/5*sin{60},5.5/5*sin{60}} \draw[thin] (\i,\j) -- +(60:1.5);
\foreach \i/\j in {1.5/5*sin{60},4.5/3*sin{60},7.5/5*sin{60},1.5/1*sin{60},4.5/-1*sin{60},7.5/1*sin{60}} \draw (\i,\j) node {\color{blue}4};
\foreach \i/\j in {1.5/3*sin{60},4.5/1*sin{60},7.5/3*sin{60},1.5/-1*sin{60},4.5/5*sin{60},7.5/-1*sin{60}} \draw (\i,\j) node {\color{blue}1};
\foreach \i/\j in {1/0,4/2*sin{60},7/0,1/4*sin{60},4/6*sin{60},7/4*sin{60}} 	
{
	\draw (\i,\j) ++(-0.7,-0.2*sin{60}) node {\color{blue}2};
	\draw (\i,\j) ++(-1.3,0.2*sin{60}) node {\color{blue}6};
}
\foreach \i/\j in {1/2*sin{60},4/0*sin{60},7/2*sin{60},1/6*sin{60},4/4*sin{60},7/6*sin{60}} 	
{
	\draw (\i,\j) ++(-0.7,-0.2*sin{60}) node {\color{blue}5};
	\draw (\i,\j) ++(-1.3,0.2*sin{60}) node {\color{blue}3};
}
\draw[color=red] (0.8,0.1) -- +(3,2*sin{60}) -- +(3,6*sin{60}) -- +(0,4*sin{60}) -- +(0,0);
\end{tikzpicture}
\quad	
\begin{tikzpicture}[scale=1.5, font = \small]
	\node[shape=circle,draw=black] (A) at (0,1.73) {\color{blue}1};
	\node[shape=circle,draw=black] (B) at (1,1.73) {\color{blue}2};
	\node[shape=circle,draw=black] (C) at (1.5,0.865) {\color{blue}3};
	\node[shape=circle,draw=black] (D) at (1,0) {\color{blue}4};
	\node[shape=circle,draw=black] (E) at (0,0) {\color{blue}5};
	\node[shape=circle,draw=black] (F) at (-0.5,0.865) {\color{blue}6};
	\path
	\foreach \from/\to in {A/B,B/C,C/D,D/E,E/F,F/A,A/C,C/E,E/A,B/D,D/F,F/B}
	{
		(\from) edge[mid arrow] (\to)					
	}
	;				
\end{tikzpicture} 		
\end{center}

\paragraph{Polygon 6c}
\begin{center}
\begin{tikzpicture}[scale=1, font = \small]
	\draw[fill] (0,1) circle (1pt) -- (1,1) circle (1pt) -- (1,0) circle (1pt) -- (0,-1) circle (1pt) -- (-1,0) circle (1pt) -- (-1,1) circle (1pt)--  (0,1);
	\draw[fill] (0,0) circle (1pt);
\end{tikzpicture}
\quad
\begin{tikzpicture}[scale=0.6, font = \small]
\foreach \i/\j in {-3/6,-2/4,-2/5,-2/6,-1/2,-1/3,-1/4,-1/5,-1/6,0/0,0/1,0/2,0/3,0/4,0/5,1/0,1/1,1/2,1/3,2/0,2/1}
{
	\coordinate (period) at (3*\i+1.5*\j,sin{60}*\j);	
	\foreach \pointb in {(1,0), (-0.5,-sin{60})}
	{
		\draw[fill] \pointb ++(period) circle (3pt);
		\foreach \a in {0,120,-120} \draw[thin] \pointb ++(period) -- +(\a:0.5);
	}
	\foreach \pointw in {(0.5,-sin{60}),  (-1,0)}
	{
		\draw \pointw ++(period) circle (3pt);
		\foreach \a in {60,180,-60} \draw[thin] \pointw ++(period) -- +(\a:0.5);
	}
}
\foreach \i/\j in {-0.5/3*sin{60},-0.5/-sin{60},2.5/3*sin{60},2.5/-sin{60},5.5/3*sin{60},5.5/-sin{60}} \draw[thin] (\i,\j) -- +(60:2);
\foreach \i/\j in {-0.5/3*sin{60},2.5/3*sin{60},5.5/3*sin{60}} \draw[thin] (\i,\j) -- +(300:2);
\foreach \i/\j in {0.5/5*sin{60},3.5/5*sin{60},6.5/5*sin{60}} \draw[thin] (\i,\j) -- +(120:1.5);
\foreach \i/\j in {1.5/5*sin{60},4.5/5*sin{60},7.5/5*sin{60},1.5/1*sin{60},4.5/1*sin{60},7.5/1*sin{60}} \draw (\i,\j) node {\color{blue}4};
\foreach \i/\j in {1.5/3*sin{60},4.5/3*sin{60},7.5/3*sin{60},1.5/-1*sin{60},4.5/-1*sin{60},7.5/-1*sin{60}} \draw (\i,\j) node {\color{blue}1};
\foreach \i/\j in {1/0,4/0,7/0,1/4*sin{60},4/4*sin{60},7/4*sin{60}} 	
{
	\draw (\i,\j) ++(-0.7,-0.2*sin{60}) node {\color{blue}3};
	\draw (\i,\j) ++(-1.3,0.2*sin{60}) node {\color{blue}2};
	\draw (\i,\j) ++(-1.3,1.8*sin{60}) node {\color{blue}6};
	\draw (\i,\j) ++(-0.7,2.3*sin{60}) node {\color{blue}5};	
}
\draw[color=red] (0,0.2) -- +(3,0) -- +(3,4*sin{60}) -- +(0,4*sin{60}) -- +(0,0);
\end{tikzpicture}
\quad	
\begin{tikzpicture}[scale=1.5, font = \small]
	\node[shape=circle,draw=black] (A) at (0,1.73) {\color{blue}1};
	\node[shape=circle,draw=black] (B) at (1,1.73) {\color{blue}2};
	\node[shape=circle,draw=black] (C) at (1.5,0.865) {\color{blue}3};
	\node[shape=circle,draw=black] (D) at (1,0) {\color{blue}4};
	\node[shape=circle,draw=black] (E) at (0,0) {\color{blue}5};
	\node[shape=circle,draw=black] (F) at (-0.5,0.865) {\color{blue}6};
	\path
	\foreach \from/\to in {A/B,B/C,C/D,D/E,E/F,F/A,A/C,C/E,E/A,B/D,D/F,F/B}
	{
		(\from) edge[mid arrow] (\to)					
	}
	;				
\end{tikzpicture} 	
\end{center}

\paragraph{Polygon 6d}
\begin{center}
\begin{tikzpicture}[scale=1, font = \small]
	\draw[fill] (0,1) circle (1pt) -- (1,0) circle (1pt) -- (-1,-1) circle (1pt) -- (-1,0) circle (1pt) -- (-1,1) circle (1pt) -- (-1,2) circle (1pt)  -- (0,1);
	\draw[fill] (0,0) circle (1pt);
\end{tikzpicture}
\quad
\begin{tikzpicture}[scale=0.5, font = \small]
\foreach \i/\j in {-3/6,-3/7,-2/4,-2/5,-2/6,-2/7,-1/2,-1/3,-1/4,-1/5,-1/6,-1/7,0/0,0/1,0/2,0/3,0/4,0/5,1/0,1/1,1/2,1/3,2/0,2/1}
{
	\coordinate (period) at (3*\i+1.5*\j,sin{60}*\j);	
	\foreach \pointb in {(1,0), (-0.5,-sin{60})}
	{
		\draw[fill] \pointb ++(period) circle (3pt);
		\foreach \a in {0,120,-120} \draw[thin] \pointb ++(period) -- +(\a:0.5);
	}
	\foreach \pointw in {(0.5,-sin{60}),  (-1,0)}
	{
		\draw \pointw ++(period) circle (3pt);
		\foreach \a in {60,180,-60} \draw[thin] \pointw ++(period) -- +(\a:0.5);
	}
	\foreach \point in {(0,0),(0,4*sin{60}),(4.5,sin{60}),(4.5,5*sin{60})} {\node at \point {\color{blue}1};}
	\foreach \point in {(0,2*sin{60}),(0,6*sin{60}),(4.5,-sin{60}),(4.5,3*sin{60}),(4.5,7*sin{60})} {\node at \point {\color{blue}4};}	
	\foreach \point in {(1.5,sin{60}),(1.5,5*sin{60}),(6,2*sin{60}),(6,6*sin{60})} {\node at \point {\color{blue}2};}
	\foreach \point in {(1.5,-sin{60}),(1.5,3*sin{60}),(1.5,7*sin{60}),(6,4*sin{60}),(6,0*sin{60})} {\node at \point {\color{blue}5};}
	\foreach \point in {(3,2*sin{60}),(3,6*sin{60}),(7.5,3*sin{60}),(7.5,7*sin{60})} {\node at \point {\color{blue}3};}	
	\foreach \point in {(3,4*sin{60}),(3,0*sin{60}),(7.5,1*sin{60}),(7.5,5*sin{60})} {\node at \point {\color{blue}6};}		
}
\draw[color=red] (-0.3,-0.3) -- +(4.5,1*sin{60}) -- +(4.5,5*sin{60}) -- +(0,4*sin{60}) -- +(0,0);
\end{tikzpicture}
\quad
\begin{tikzpicture}[scale=1.5, font = \small]
	\node[shape=circle,draw=black] (A) at (0,1.73) {\color{blue}1};
	\node[shape=circle,draw=black] (B) at (1,1.73) {\color{blue}2};
	\node[shape=circle,draw=black] (C) at (1.5,0.865) {\color{blue}3};
	\node[shape=circle,draw=black] (D) at (1,0) {\color{blue}4};
	\node[shape=circle,draw=black] (E) at (0,0) {\color{blue}5};
	\node[shape=circle,draw=black] (F) at (-0.5,0.865) {\color{blue}6};
	\path
	\foreach \from/\to in {A/B,B/C,C/D,D/E,E/F,F/A,A/C,C/E,E/A,B/D,D/F,F/B}
	{
		(\from) edge[mid arrow] (\to)					
	}
	;				
\end{tikzpicture} 	
\end{center}

\paragraph{Polygon 7a}
\begin{center}
\begin{tikzpicture}[scale=1, font = \small]
	\draw[fill] (0,1) circle (1pt) -- (1,0) circle (1pt) -- (0,-1) circle (1pt) -- (-1,-1) circle (1pt) -- (-1,0) circle (1pt) -- (-1,1) circle (1pt) -- (-1,2) circle (1pt)  -- (0,1);
	\draw[fill] (0,0) circle (1pt);
\end{tikzpicture}
\quad
\begin{tikzpicture}[scale=0.6, font = \small]
\foreach \i/\j in {-3/6,-3/7,-2/4,-2/5,-2/6,-2/7,-1/2,-1/3,-1/4,-1/5,-1/6,-1/7,0/0,0/1,0/2,0/3,0/4,0/5,1/0,1/1,1/2,1/3,2/0,2/1}
{
	\coordinate (period) at (3*\i+1.5*\j,sin{60}*\j);	
	\foreach \pointb in {(1,0), (-0.5,-sin{60})}
	{
		\draw[fill] \pointb ++(period) circle (2.5pt);
		\foreach \a in {0,120,-120} \draw[thin] \pointb ++(period) -- +(\a:0.5);
	}
	\foreach \pointw in {(0.5,-sin{60}),  (-1,0)}
	{
		\draw \pointw ++(period) circle (2.5pt);
		\foreach \a in {60,180,-60} \draw[thin] \pointw ++(period) -- +(\a:0.5);
	}
}
	\foreach \point in {(1,0),(1,4*sin{60}),(5.5,-sin{60}),(5.5,3*sin{60})} {\draw[thin] \point -- +(60:2);}
	\draw[thin] (5.5,7*sin{60}) --+(60:0.5);
	\foreach \point in {(1.5,-sin{60}),(1.5,3*sin{60}),(1.5,7*sin{60}),(6,6*sin{60}),(6,2*sin{60})} {\node at \point {\color{blue}6};}
	\foreach \point in {(3,0*sin{60}),(3,4*sin{60}),(7.5,7*sin{60}),(7.5,3*sin{60}),(7.5,-1*sin{60})} {\node at \point {\color{blue}2};}
	\foreach \point in {(0,0*sin{60}),(0,4*sin{60}),(4.5,3*sin{60}),(4.	5,-1*sin{60})} {\node at \point {\color{blue}7};}
	\foreach \point in {(0,2*sin{60}),(0,6*sin{60}),(4.5,7*sin{60}),(4.5,5*sin{60}),(4.5,1*sin{60})} {\node at \point {\color{blue}4};}
	\foreach \point in {(3,2*sin{60}),(3,6*sin{60}),(7.5,5*sin{60}),(7.5,sin{60})} {\node at \point {\color{blue}5};}
	\foreach \point in {(1.9,0.8*sin{60}),(1.9,4.8*sin{60}),(6.3,3.8*sin{60}),(6.3,-0.2*sin{60})} {\node at \point {\color{blue}3};}
	\foreach \point in {(1.2,1.2*sin{60}),(1.2,5.2*sin{60}),(5.7,4.2*sin{60}),(5.7,0.2*sin{60})} {\node at \point {\color{blue}1};}
\draw[color=red] (-0,-0) -- +(4.5,-1*sin{60}) -- +(4.5,3*sin{60}) -- +(0,4*sin{60}) -- +(0,0);
\end{tikzpicture}
\quad
\begin{tikzpicture}[scale=2, font = \small]
	\node[shape=circle,draw=black] (A) at (0,1) {\color{blue}1};
	\node[shape=circle,draw=black] (B) at (-0.4,1.4) {\color{blue}2};
	\node[shape=circle,draw=black] (C) at (1,1) {\color{blue}3};
	\node[shape=circle,draw=black] (D) at (1.4,1.4) {\color{blue}4};
	\node[shape=circle,draw=black] (E) at (1,-0.3) {\color{blue}5};
	\node[shape=circle,draw=black] (F) at (0.5,0.0) {\color{blue}6};
	\node[shape=circle,draw=black] (G) at (0,-0.3) {\color{blue}7};
	\path
	\foreach \from/\to in {B/D,D/E,E/G,G/A,D/F,G/B,F/B}
	{
		(\from) edge[mid arrow, bend left=20] (\to)					
	}
	\foreach \from/\to in {A/D,B/C,E/F,F/G}
	{
		(\from) edge[mid arrow] (\to)					
	}
	\foreach \from/\to in {A/C,C/E,C/F,F/A}
	{
		(\from) edge[mid arrow, bend right=20] (\to)					
	}
	;		
\end{tikzpicture}

\end{center}

\paragraph{Polygon 7b}
\begin{center}
\begin{tikzpicture}[scale=1, font = \small]
	\draw[fill] (0,1) circle (1pt) -- (1,0) circle (1pt) -- (1,-1) circle (1pt) -- (0,-1) circle (1pt) -- (-1,-1) circle (1pt) -- (-1,0) circle (1pt) -- (-1,1) circle (1pt) --  (0,1);
	\draw[fill] (0,0) circle (1pt);
\end{tikzpicture}
\quad
\begin{tikzpicture}[scale=0.6, font = \small]
\foreach \i/\j in {-3/6,-2/4,-2/5,-2/6,-1/2,-1/3,-1/4,-1/5,-1/6,0/0,0/1,0/2,0/3,0/4,0/5,1/0,1/1,1/2,1/3,2/0,2/1}
{
	\coordinate (period) at (3*\i+1.5*\j,sin{60}*\j);	
	\foreach \pointb in {(1,0), (-0.5,-sin{60})}
	{
		\draw[fill] \pointb ++(period) circle (3pt);
		\foreach \a in {0,120,-120} \draw[thin] \pointb ++(period) -- +(\a:0.5);
	}
	\foreach \pointw in {(0.5,-sin{60}),  (-1,0)}
	{
		\draw \pointw ++(period) circle (3pt);
		\foreach \a in {60,180,-60} \draw[thin] \pointw ++(period) -- +(\a:0.5);
	}
}
\foreach \i/\j in {-1/2*sin{60},-1/6*sin{60},0.5/1*sin{60},0.5/-sin{60},0.5/3*sin{60},0.5/5*sin{60},2/0*sin{60},2/4*sin{60},3.5/1*sin{60},3.5/-sin{60},3.5/3*sin{60},3.5/5*sin{60},5/2*sin{60},5/6*sin{60},6.5/1*sin{60},6.5/3*sin{60},6.5/5*sin{60}} \draw[thin] (\i,\j) -- +(0:2);
\foreach \i/\j in {6.5/-1*sin{60},8/0*sin{60},8/2*sin{60},8/4*sin{60}} \draw[thin] (\i,\j) -- +(0:0.5);
\foreach \i/\j in {-0.5/-1*sin{60},-0.5/1*sin{60},-0.5/3*sin{60},-0.5/5*sin{60}} \draw[thin] (\i,\j) -- +(180:0.5);
\foreach \i/\j in {0/0,0/4*sin{60},3/2*sin{60},3/6*sin{60},6/0*sin{60},6/4*sin{60}} \draw (\i,\j) node {\color{blue}6};
\foreach \i/\j in {0/2*sin{60},0/6*sin{60},3/0*sin{60},3/4*sin{60},6/2*sin{60},6/6*sin{60}}
	\draw (\i,\j) ++(0,-0.5*sin{60}) node {\color{blue}5};
\foreach \i/\j in {0/2*sin{60},3/0*sin{60},3/4*sin{60},6/2*sin{60}}
	\draw (\i,\j) ++(0,0.5*sin{60}) node {\color{blue}7};
\foreach \i/\j in {1.5/1*sin{60},1.5/5*sin{60},4.5/-1*sin{60},4.5/3*sin{60},7.5/1*sin{60},7.5/5*sin{60}}
\draw (\i,\j) ++(0,0.5*sin{60}) node {\color{blue}3};
\foreach \i/\j in {1.5/1*sin{60},1.5/5*sin{60},4.5/3*sin{60},7.5/1*sin{60},7.5/5*sin{60}}
\draw (\i,\j) ++(0,-0.5*sin{60}) node {\color{blue}1};
\foreach \i/\j in {1.5/-1*sin{60},1.5/3*sin{60},4.5/1*sin{60},4.5/5*sin{60},7.5/-1*sin{60},7.5/3*sin{60}}
\draw (\i,\j) ++(0,0.5*sin{60}) node {\color{blue}4};
\foreach \i/\j in {1.5/3*sin{60},4.5/1*sin{60},4.5/5*sin{60},7.5/3*sin{60}}
\draw (\i,\j) ++(0,-0.5*sin{60}) node {\color{blue}2};
\draw[color=red] (0,0) -- +(3,2*sin{60}) -- +(3,6*sin{60}) -- +(0,4*sin{60}) -- +(0,0);
\end{tikzpicture}
\quad
\begin{tikzpicture}[scale=2, font = \small]
	\node[shape=circle,draw=black] (A) at (0,1) {\color{blue}1};
	\node[shape=circle,draw=black] (B) at (-0.4,1.4) {\color{blue}2};
	\node[shape=circle,draw=black] (C) at (1,1) {\color{blue}3};
	\node[shape=circle,draw=black] (D) at (1.4,1.4) {\color{blue}4};
	\node[shape=circle,draw=black] (E) at (1,-0.3) {\color{blue}5};
	\node[shape=circle,draw=black] (F) at (0.5,0.0) {\color{blue}6};
	\node[shape=circle,draw=black] (G) at (0,-0.3) {\color{blue}7};
	\path
	\foreach \from/\to in {B/D,D/E,E/G,G/A,D/F,G/B,F/B}
	{
		(\from) edge[mid arrow, bend left=20] (\to)					
	}
	\foreach \from/\to in {A/D,B/C,E/F,F/G}
	{
		(\from) edge[mid arrow] (\to)					
	}
	\foreach \from/\to in {A/C,C/E,C/F,F/A}
	{
		(\from) edge[mid arrow, bend right=20] (\to)					
	}
	;		
\end{tikzpicture}
\end{center}

\paragraph{Polygon 8a}
\begin{center}
\begin{tikzpicture}[scale=1, font = \small]
	\draw[fill] (0,1) circle (1pt) -- (1,-1) circle (1pt) -- (0,-1) circle (1pt) -- (-1,-1) circle (1pt) -- (-1,0) circle (1pt) -- (-1,1) circle (1pt) -- (-1,2) circle (1pt) -- (-1,3) circle (1pt) -- (0,1);
	\draw[fill] (0,0) circle (1pt);	
\end{tikzpicture}
\quad
\begin{tikzpicture}[scale=0.5, font = \small]
\foreach \i/\j in {-3/6,-3/7,-2/4,-2/5,-2/6,-2/7,-1/2,-1/3,-1/4,-1/5,-1/6,-1/7,0/0,0/1,0/2,0/3,0/4,0/5,1/0,1/1,1/2,1/3,2/0,2/1}
{
	\coordinate (period) at (3*\i+1.5*\j,sin{60}*\j);	
	\foreach \pointb in {(1,0), (-0.5,-sin{60})}
	{
		\draw[fill] \pointb ++(period) circle (3pt);
		\foreach \a in {0,120,-120} \draw[thin] \pointb ++(period) -- +(\a:0.5);
	}
	\foreach \pointw in {(0.5,-sin{60}),  (-1,0)}
	{
		\draw \pointw ++(period) circle (3pt);
		\foreach \a in {60,180,-60} \draw[thin] \pointw ++(period) -- +(\a:0.5);
	}
	\foreach \point in {(0,0),(0,4*sin{60}),(6,0),(6,4*sin{60})} {\node at \point {\color{blue}1};}
	\foreach \point in {(3,0),(3,4*sin{60})} {\node at \point {\color{blue}2};}	
	\foreach \point in {(0,2*sin{60}),(0,6*sin{60}),(6,2*sin{60}),(6,6*sin{60})} {\node at \point {\color{blue}5};}
	\foreach \point in {(3,2*sin{60}),(3,6*sin{60})} {\node at \point {\color{blue}6};}	
	\foreach \point in {(1.5,sin{60}),(1.5,5*sin{60}),(7.5,sin{60}),(7.5,5*sin{60})} {\node at \point {\color{blue}3};}	
	\foreach \point in {(4.5,sin{60}),(4.5,5*sin{60})} {\node at \point {\color{blue}4};}		
	\foreach \point in {(1.5,-sin{60}),(1.5,3*sin{60}),(1.5,7*sin{60}),(7.5,3*sin{60}),(7.5,7*sin{60})} {\node at \point {\color{blue}7};}	
	\foreach \point in {(4.5,3*sin{60}),(4.5,-sin{60}),(4.5,7*sin{60})} {\node at \point {\color{blue}8};}
}
\draw[color=red] (0.3,1.3) -- +(6,0) -- +(6,4*sin{60}) -- +(0,4*sin{60}) -- +(0,0);
\end{tikzpicture}
\quad
\begin{tikzpicture}[scale=2, font = \small]
	\node[shape=circle,draw=black] (A) at (0,1) {\color{blue}1};
	\node[shape=circle,draw=black] (A1) at(-0.3,1.3) {\color{blue}2};
	\node[shape=circle,draw=black] (B) at (1,1) {\color{blue}3};
	\node[shape=circle,draw=black] (B1) at(1.3,1.3) {\color{blue}4};
	\node[shape=circle,draw=black] (C) at (1,0) {\color{blue}5};
	\node[shape=circle,draw=black] (C1) at(1.3,-0.3) {\color{blue}6};
	\node[shape=circle,draw=black] (D) at (0,0) {\color{blue}7};
	\node[shape=circle,draw=black] (D1) at(-0.3,-0.3) {\color{blue}8};
	\path
	\foreach \from/\to in {A/B1,A1/B,B/C1,B1/C,C/D1,C1/D,D/A1,D1/A}
	{
		(\from) edge[mid arrow] (\to)					
	}
	\foreach \from/\to in {A/B,B/C,C/D,D/A}
	{
		(\from) edge[mid arrow,bend right=15] (\to)					
	}
	\foreach \from/\to in {A1/B1,B1/C1,C1/D1,D1/A1}
	{
		(\from) edge[mid arrow,bend left=15] (\to)					
	}
	;
\end{tikzpicture}
\end{center}

\paragraph{Polygon 8b}
\begin{center}
\begin{tikzpicture}[scale=1, font = \small]
\draw[fill] (0,1) circle (1pt) -- (1,0) circle (1pt) -- (1,-1) circle (1pt) -- (0,-1) circle (1pt) -- (-1,-1) circle (1pt) -- (-1,0) circle (1pt) -- (-1,1) circle (1pt) -- (-1,2) circle (1pt) -- (0,1);
\draw[fill] (0,0) circle (1pt);	
\end{tikzpicture}
\quad
\begin{tikzpicture}[scale=0.6, font = \small]
\foreach \i/\j in {-3/6,-2/4,-2/5,-2/6,-1/2,-1/3,-1/4,-1/5,-1/6,0/0,0/1,0/2,0/3,0/4,0/5,1/0,1/1,1/2,1/3,2/0,2/1}
{
	\coordinate (period) at (3*\i+1.5*\j,sin{60}*\j);	
	\foreach \pointb in {(1,0), (-0.5,-sin{60})}
	{
		\draw[fill] \pointb ++(period) circle (3pt);
		\foreach \a in {0,120,-120} \draw[thin] \pointb ++(period) -- +(\a:0.5);
	}
	\foreach \pointw in {(0.5,-sin{60}),  (-1,0)}
	{
		\draw \pointw ++(period) circle (3pt);
		\foreach \a in {60,180,-60} \draw[thin] \pointw ++(period) -- +(\a:0.5);
	}
}
\foreach \i/\j in {1/0*sin{60},1/2*sin{60},1/4*sin{60},5.5/3*sin{60},5.5/-sin{60},5.5/1*sin{60}} \draw[thin] (\i,\j) -- +(60:2);
\foreach \i/\j in {5.5/5*sin{60}} \draw[thin] (\i,\j) -- +(60:1.5);
\foreach \i/\j in {2/0*sin{60}} \draw[thin] (\i,\j) -- +(240:1.5);
\foreach \i/\j in {1/6*sin{60}} \draw[thin] (\i,\j) -- +(60:0.5);
\foreach \i/\j in {6.5/-1*sin{60}} \draw[thin] (\i,\j) -- +(240:0.5);
\foreach \i/\j in {0/2*sin{60},0/6*sin{60},4.5/1*sin{60},4.5/5*sin{60}} \draw (\i,\j) node {\color{blue}1};
\foreach \i/\j in {0/0*sin{60},0/4*sin{60},4.5/-1*sin{60},4.5/3*sin{60}} \draw (\i,\j) node {\color{blue}5};
\foreach \i/\j in {3/0,3/4*sin{60},7.5/-1*sin{60},7.5/3*sin{60}} \draw (\i,\j) node {\color{blue}7};
\foreach \i/\j in {3/2*sin{60},3/6*sin{60},7.5/1*sin{60},7.5/5*sin{60}} \draw (\i,\j) node {\color{blue}3};
\foreach \i/\j in {2.5/sin{60},2.5/5*sin{60},7/0,7/4*sin{60}} 	
{
	\draw (\i,\j) ++(-0.7,-0.2*sin{60}) node {\color{blue}2};
	\draw (\i,\j) ++(-1.3,0.2*sin{60}) node {\color{blue}8};
}
\foreach \i/\j in {2.5/-sin{60},2.5/3*sin{60},7/2*sin{60},7/6*sin{60}} 	
{
	\draw (\i,\j) ++(-0.7,-0.2*sin{60}) node {\color{blue}6};
	\draw (\i,\j) ++(-1.3,0.2*sin{60}) node {\color{blue}4};
}
\draw[color=red] (0,0.1) -- +(4.5,-sin{60}) -- +(4.5,3*sin{60}) -- +(0,4*sin{60}) -- +(0,0);
\end{tikzpicture}
\quad
\begin{tikzpicture}[scale=2, font = \small]
	\node[shape=circle,draw=black] (A) at (0,1) {\color{blue}1};
	\node[shape=circle,draw=black] (A1) at(-0.3,1.3) {\color{blue}2};
	\node[shape=circle,draw=black] (B) at (1,1) {\color{blue}3};
	\node[shape=circle,draw=black] (B1) at(1.3,1.3) {\color{blue}4};
	\node[shape=circle,draw=black] (C) at (1,0) {\color{blue}5};
	\node[shape=circle,draw=black] (C1) at(1.3,-0.3) {\color{blue}6};
	\node[shape=circle,draw=black] (D) at (0,0) {\color{blue}7};
	\node[shape=circle,draw=black] (D1) at(-0.3,-0.3) {\color{blue}8};
	\path
	\foreach \from/\to in {A/B1,A1/B,B/C1,B1/C,C/D1,C1/D,D/A1,D1/A}
	{
		(\from) edge[mid arrow] (\to)					
	}
	\foreach \from/\to in {A/B,B/C,C/D,D/A}
	{
		(\from) edge[mid arrow,bend right=15] (\to)					
	}
	\foreach \from/\to in {A1/B1,B1/C1,C1/D1,D1/A1}
	{
		(\from) edge[mid arrow,bend left=15] (\to)					
	}
	;
\end{tikzpicture}
\end{center}

\paragraph{Polygon 8c}
\begin{center}
\begin{tikzpicture}[scale=1, font = \small]
	\draw[fill] (0,1) circle (1pt) -- (1,1) circle (1pt) -- (1,0) circle (1pt) -- (1,-1) circle (1pt) -- (0,-1) circle (1pt) -- (-1,-1) circle (1pt) -- (-1,0) circle (1pt) -- (-1,1) circle (1pt) -- (0,1);
	\draw[fill] (0,0) circle (1pt);	
\end{tikzpicture}
\quad
\begin{tikzpicture}[scale=0.8, font = \small]
\foreach \i/\j in {-1/1,0/0,0/1,1/-1,1/0,1/1,2/-1,2/0,3/0,3/-1,3/-2,4/-2,4/-1,5/-2}
{
	\coordinate (period) at (\i,\i+2*\j);	
	\foreach \pointb in {(0.5,-0.5), (-0.5,0.5)}
	{
		\draw[fill] \pointb ++(period) circle (2pt);
		\foreach \a in {0,90,180,270} \draw[thin] \pointb ++(period) -- +(\a:0.5);
	}
	\foreach \pointw in {(0.5,0.5),  (-0.5,-0.5)}
	{
		\draw \pointw ++(period) circle (2pt);
		\foreach \a in {0,90,180,270} \draw[thin] \pointw ++(period) -- +(\a:0.5);
	}
	\foreach \point in {(0,2),(2,0),(4,2)} {\node at \point {\color{blue}1};}	
	\foreach \point in {(0,0),(2,2),(4,0)} {\node at \point {\color{blue}2};}	
	\foreach \point in {(0,1),(2,-1),(2,3),(4,1)} {\node at \point {\color{blue}8};}		
	\foreach \point in {(2,1),(4,3),(0,3),(4,-1),(0,-1)} {\node at \point {\color{blue}7};}			
	\foreach \point in {(-1,0),(1,2),(3,0),(5,2)} {\node at \point {\color{blue}3};}	
	\foreach \point in {(-1,2),(1,0),(3,2),(5,0)} {\node at \point {\color{blue}4};}	
	\foreach \point in {(-1,1),(1,3),(3,1),(1,-1)} {\node at \point {\color{blue}5};}	
	\foreach \point in {(3,-1),(1,1),(3,3),(5,1)} {\node at \point {\color{blue}6};}		
}
\draw[color=red] (-0,1) -- +(2,2) -- +(4,0) -- +(2,-2) -- +(0,0);
\end{tikzpicture}
\quad
\begin{tikzpicture}[scale=2, font = \small]
	\node[shape=circle,draw=black] (A) at (0,1) {\color{blue}1};
	\node[shape=circle,draw=black] (A1) at(-0.3,1.3) {\color{blue}2};
	\node[shape=circle,draw=black] (B) at (1,1) {\color{blue}3};
	\node[shape=circle,draw=black] (B1) at(1.3,1.3) {\color{blue}4};
	\node[shape=circle,draw=black] (C) at (1,0) {\color{blue}5};
	\node[shape=circle,draw=black] (C1) at(1.3,-0.3) {\color{blue}6};
	\node[shape=circle,draw=black] (D) at (0,0) {\color{blue}7};
	\node[shape=circle,draw=black] (D1) at(-0.3,-0.3) {\color{blue}8};
	\path
	\foreach \from/\to in {A/B1,A1/B,B/C1,B1/C,C/D1,C1/D,D/A1,D1/A}
	{
		(\from) edge[mid arrow] (\to)					
	}
	\foreach \from/\to in {A/B,B/C,C/D,D/A}
	{
		(\from) edge[mid arrow,bend right=15] (\to)					
	}
	\foreach \from/\to in {A1/B1,B1/C1,C1/D1,D1/A1}
	{
		(\from) edge[mid arrow,bend left=15] (\to)					
	}
	;
\end{tikzpicture}
\end{center}

\paragraph{Polygon 9}
\begin{center}
\begin{tikzpicture}[scale=1, font = \small]
\draw[fill] (0,1) circle (1pt) -- (1,0) circle (1pt) -- (2,-1) circle (1pt) -- (1,-1) circle (1pt) -- (0,-1) circle (1pt) -- (-1,-1) circle (1pt) -- (-1,0) circle (1pt) -- (-1,1) circle (1pt) -- (-1,2) circle (1pt) -- (0,1);
\draw[fill] [](0,0) circle (1pt);	
\end{tikzpicture}
\quad
\begin{tikzpicture}[scale=0.5, font = \small]
\foreach \i/\j in {-4/8,-4/9,-3/6,-3/7,-3/8,-3/9,-2/4,-2/5,-2/6,-2/7,-2/8,-2/9,-1/2,-1/3,-1/4,-1/5,-1/6,-1/7,0/0,0/1,0/2,0/3,0/4,0/5,1/0,1/1,1/2,1/3,2/0,2/1}
{
	\coordinate (period) at (3*\i+1.5*\j,sin{60}*\j);	
	\foreach \pointb in {(1,0), (-0.5,-sin{60})}
	{
		\draw[fill] \pointb ++(period) circle (3pt);
		\foreach \a in {0,120,-120} \draw[thin] \pointb ++(period) -- +(\a:0.5);
	}
	\foreach \pointw in {(0.5,-sin{60}),  (-1,0)}
	{
		\draw \pointw ++(period) circle (3pt);
		\foreach \a in {60,180,-60} \draw[thin] \pointw ++(period) -- +(\a:0.5);
	}
}
	\foreach \point in {(0,0),(0,6*sin{60}),(4.5,3*sin{60}),(4.5,9*sin{60})} {\node at \point {\color{blue}1};}
	\foreach \point in {(1.5,3*sin{60}),(1.5,9*sin{60}),(6,6*sin{60}),(6,0)} {\node at \point {\color{blue}2};}
	\foreach \point in {(3,6*sin{60}),(7.5,3*sin{60}),(7.5,9*sin{60}),(3,0)} {\node at \point {\color{blue}3};}
	\foreach \point in {(1.5,sin{60}),(1.5,7*sin{60}),(6,4*sin{60})} {\node at \point {\color{blue}4};}
	\foreach \point in {(3,4*sin{60}),(7.5,7*sin{60}),(7.5,sin{60})} {\node at \point {\color{blue}5};}
	\foreach \point in {(0,4*sin{60}),(4.5,7*sin{60}),(4.5,1*sin{60})} {\node at \point {\color{blue}6};}
	\foreach \point in {(0,2*sin{60}),(4.5,5*sin{60}),(0,8*sin{60})} {\node at \point {\color{blue}8};}
	\foreach \point in {(3,2*sin{60}),(7.5,5*sin{60}),(3,8*sin{60})} {\node at \point {\color{blue}7};}
	\foreach \point in {(3,2*sin{60}),(7.5,5*sin{60}),(3,8*sin{60})} {\node at \point {\color{blue}7};}
	\foreach \point in {(1.5,5*sin{60}),(6,8*sin{60}),(6,2*sin{60})} {\node at \point {\color{blue}9};}
\draw[color=red] (-0.3,-0.3) -- +(4.5,3*sin{60}) -- +(4.5,9*sin{60}) -- +(0,6*sin{60}) -- +(0,0);
\end{tikzpicture}
\quad
\begin{tikzpicture}[scale=2, font = \small]
	\node[shape=circle,draw=black] (A) at (-0.5,0) {\color{blue}1};
	\node[shape=circle,draw=black] (A1) at (-0.76,-0.15) {\color{blue}2};
	\node[shape=circle,draw=black] (A2) at (-1.02,-0.3) {\color{blue}3};	
	\node[shape=circle,draw=black] (B) at (0.5,0) {\color{blue}7};
	\node[shape=circle,draw=black] (B1) at (0.76,-0.15) {\color{blue}8};
	\node[shape=circle,draw=black] (B2) at (1.02,-0.30) {\color{blue}9};	
	\node[shape=circle,draw=black]  (C) at (0,0.86) {\color{blue}4};
	\node[shape=circle,draw=black] (C1) at (0,1.16) {\color{blue}5};
	\node[shape=circle,draw=black] (C2) at (0,1.46) {\color{blue}6};	
	\path
	\foreach \from/\to in {C1/B,B1/A,A1/C}
	{
		(\from) edge[mid arrow] (\to)					
	}
	\foreach \from/\to in {C2/B,C2/B1,C2/B2,B2/A,B2/A1,B2/A2,A2/C,A2/C1,A2/C2}
	{
		(\from) edge[mid arrow,bend left=30] (\to)					
	}
	\foreach \from/\to in {C/B,B/A,A/C}
	{
		(\from) edge[mid arrow,bend right=30] (\to)					
	}
	\foreach \from/\to in {C/B1,B/A1,A/C1}
	{
		(\from) edge[mid arrow,bend right=20] (\to)					
	}
	\foreach \from/\to in {C/B2,B/A2,A/C2}
	{
		(\from) edge[mid arrow,bend right=10] (\to)					
	}
	\foreach \from/\to in {C1/B2,B1/A2,A1/C2}
	{
		(\from) edge[mid arrow,bend left=15] (\to)					
	}
	\foreach \from/\to in {C1/B1,B1/A1,A1/C1}
	{
		(\from) edge[mid arrow,bend left=8] (\to)					
	}
	;
\end{tikzpicture}
\end{center}
\section{$q$-special function and $q$-conformal blocks} \label{App:q-functions}
Infinite multiple $q$-deformed Pochhammer symbol is defined by
\begin{equation}
(x;t_1,\ldots t_N)_{\infty}=\prod_{i_1,\ldots i_N=0}^{\infty}\left(1-x\prod_{k=1}^Nt_k^{i_k}\right)
=\exp\left(-\sum_{m=1}^{\infty}\frac{x^m}m\prod_{k=1}^N\frac1{1-t_k^m}\right).
\end{equation}
The product exists if all $|t_k|<1$. The exponent has meaning in larger region, and the function $(x;t_1,\ldots t_N)_{\infty}$ defined by the second expression satisfies
\begin{equation}
(x;t_1^{-1},t_2,\ldots t_N)_{\infty}=(xt_1;t_1,\ldots t_N)^{-1}_{\infty} \label{qtrans}
\end{equation}

By $\mathcal F(u ,q_1^2,q_1^{-1}q_2|Z)$ we denote (the Whittaker limit) of conformal block of $q$-deformed Virasoro algebra. The standard definition uses just a norm of the Whittaker vector. Due to AGT relation (proposed in this case in \cite{AY} and proven in \cite{Yanagida}) this function is equal to the Nekrasov instanton partition function for pure $SU(2)$ 5d gauge theory, and we use the latter as a definition

\begin{equation} \label{eq:confblock}
\mathcal F(u_1,u_2;q_1,q_2|Z)=\sum_{\lambda_1,\lambda_2} Z^{|\lambda_1|+|\lambda_2|}\frac{1}{\prod_{i,j=1}^2 N_{\lambda_i,\lambda_j}(u_i/u_j;q_1,q_2)},
\end{equation}
where
\begin{equation}
N_{\lambda,\mu}(u,q_1,q_2)
=\prod_{s\in \lambda}
(1-u q_2^{-a_\mu(s)-1}q_1^{\ell_\lambda(s)}) \cdot
\prod_{s \in \mu}
(1-u q_2^{a_\lambda(s)}q_1^{-\ell_\mu(s)-1}).\label{Nlm}
\end{equation}
The sum in \eqref{eq:confblock} runs over all pairs of Young diagrams $\lambda_1, \lambda_2$. In formula \eqref{Nlm} $a_{\lambda}(s), l_{\lambda}(s)$ are lengths of arms and legs for the box $s$ in diagram $\lambda$.

The function $\mathcal{F}$ depends on $u_1,u_2$ through their ratio $u=u_1/u_2$, so we use just $u$-variable. Parameterizing $u=e^{2Ra}, q_1=e^{R\epsilon_1}, q_2=e^{R\epsilon_2}$, in terms of conformal field theory one gets $a$ as momentum of the Virasoro representation, while $\epsilon_1,\epsilon_2$ parameterize the central charge. In terms of gauge theory $a$ is vacuum expectation value of the scalar field, $\epsilon_1, \epsilon_2$ are parameters of $\Omega$-deformation, and $R$ is a radius of the 5th compact dimension.

The convergence of the series \eqref{eq:confblock} was proven in several cases, in \cite{BS:2016:1} for $q_1q_2=1$ and infinite radius of convergence,  in \cite{Felder:2017} for $q_1,q_2>1$ or $q_2=\bar{q_1}$, $|q_1|>1$ and nonzero radius of convergence.

In the main text we exploit the function $\mathcal{F}$ in the formulas for tau-functions and in bilinear relations. In order to write them in compact form we introduce notations
\be
\label{CFdef}
C_q(u;q_1,q_2)=(u;q_1,q_2)_\infty(u^{-1};q_1,q_2)_\infty,\quad c_q(u|Z)=\exp\left(\frac{-\log Z\left(\log u \right)^2}{4\log q_1\log q_2}\right),\\
\mathrm F(u;q_1,q_2|Z)=C_q(u;q_1,q_2)\mathcal F(u;q_1,q_2|Z), \quad \quad \mathsf{F}(u;q_1,q_2|Z)=c_q(u|Z)  \mathrm F(u;q_1,q_2|Z)\\
\mathrm F^{(1)}(u|Z)=\mathrm F(u;q_1^2,q_1^{-1}q_2|Z),\quad \mathrm F^{(2)}(u|z)=\mathrm F(u;q_1q_2^{-1},q_2^{2}|Z)
\\
\mathsf F^{(1)}(u|Z)=\mathsf F(u;q_1^2,q_1^{-1}q_2|Z),\quad \mathsf F^{(2)}(u|Z)=\mathsf F(u;q_1q_2^{-1},q_2^{2}|Z)
\ee
The function $\mathsf F$ should be understood as a good normalization of $\mathcal{F}$. The another function $\mathrm{F}$ is used only in bilinear relations \eqref{eq:FT1T3}, \eqref{eq:FT1T4}, \eqref{eq:FT1T2}, \eqref{eq:FT1T1}, where we used $\mathrm{F}$ in order to stress their algebraic nature as series in $Z$.


\footnotesize

\bigskip
\noindent \textsc{Landau Institute for Theoretical Physics, Chernogolovka, Russia,\\
	Center for Advanced Studies, Skoltech, Moscow, Russia,\\
	Laboratory for Mathematical Physics, NRU HSE, Moscow, Russia,\\
	Institute for Information Transmission Problems, Moscow, Russia,\\
	Independent University of Moscow, Moscow, Russia}

\emph{E-mail}:\,\,\textbf{mbersht@gmail.com}\\

\noindent \textsc{Center for Advanced Studies, Skoltech, Moscow, Russia,\\
	Department of Mathematics and Laboratory for Mathematical Physics, NRU HSE, Moscow, Russia,\\
	Bogolyubov Institute for Theoretical Physics, Kyiv, Ukraine}

\emph{E-mail}:\,\,\textbf{pasha145@gmail.com}\\

\noindent \textsc{Center for Advanced Studies, Skoltech, Moscow, Russia,\\
	Department of Mathematics and Laboratory for Mathematical Physics, NRU HSE, Moscow, Russia,\\
	Institute for Theoretical and Experimental Physics, Moscow, Russia,\\
	Theory Department of Lebedev Physics Institute, Moscow, Russia}

\emph{E-mail}:\,\,\textbf{andrei.marshakov@gmail.com}

\end{document}